\documentclass[12pt,
 reprint,
superscriptaddress,
 amsmath,amssymb,
 aps,
 showkeys, 
]{revtex4-2}

\usepackage{graphicx}
\usepackage{dcolumn}
\usepackage{listings, tabularx}
\usepackage{amsmath,amsthm, bbm}
\usepackage[export]{adjustbox}
\usepackage{bm}

\newcommand{\bra}[1]{\ensuremath{\left\langle#1\right|}}
\newcommand{\ket}[1]{\ensuremath{\left|#1\right\rangle}}

\newcommand{\braket}[2]{\ensuremath{\left\langle#1|#2\right\rangle}}
\newcommand{\ketbra}[2]{\ensuremath{\left|#1\rangle\langle#2\right|}}

\theoremstyle{definition}

\usepackage{tikz}
\usepackage{xcolor}

\definecolor{codegreen}{rgb}{0,0.6,0}
\definecolor{codegray}{rgb}{0.5,0.5,0.5}
\definecolor{codepurple}{rgb}{0.58,0,0.82}
\definecolor{backcolour}{rgb}{0.95,0.95,0.92}

\lstdefinestyle{mystyle}{
    backgroundcolor=\color{backcolour},   
    commentstyle=\color{codegreen},
    keywordstyle=\color{magenta},
    numberstyle=\tiny\color{codegray},
    stringstyle=\color{codepurple},
    basicstyle=\ttfamily\footnotesize,
    breakatwhitespace=false,         
    breaklines=true,                 
    captionpos=b,                    
    keepspaces=true,                 
    numbers=left,                    
    numbersep=5pt,                  
    showspaces=false,                
    showstringspaces=false,
    showtabs=false,                  
    tabsize=2
}

\begin{document}

\preprint{APS/123-QED}

\title{Quantum walks as thermalizations, with application to fullerene graphs\\}

\author{Shyam Dhamapurkar}%
 \email{11930957@mail.sustech.edu.cn}
\affiliation{Shenzhen Institute for Quantum Science and Engineering and Department of Physics, Southern University of Science and Technology, Shenzhen 518055, China}

\author{Oscar Dahlsten}
\affiliation{Department of Physics, City University of Hong Kong, Tat Chee Avenue, Kowloon, Hong Kong SAR, China}
\affiliation{Shenzhen Institute for Quantum Science and Engineering and Department of Physics, Southern University of Science and Technology, Shenzhen 518055, China}
\affiliation{Institute of Nanoscience and Applications, Southern University of Science and Technology, Shenzhen 518055, China}



\date{\today}

\begin{abstract} We consider to what extent quantum walks can constitute models of thermalization, analogously to how classical random walks can be models for classical thermalization. In a quantum walk over a graph, a walker moves in a superposition of node positions via a unitary time evolution. We show a quantum walk can be interpreted as an equilibration of a kind investigated in the literature on thermalization in unitarily evolving quantum systems. This connection implies that recent results concerning the equilibration of observables can be applied to analyse the node position statistics of quantum walks. We illustrate this in the case of a family of graphs known as fullerenes. We find that a bound from Short et al., implying that certain expectation values will at most times be close to their time-averaged value, applies tightly to the node position probabilities. Nevertheless, the node position statistics do not thermalize in the standard sense. In particular, quantum walks over fullerene graphs constitute a counter-example to the hypothesis that subsystems equilibrate to the Gibbs state. We also exploit the bridge created to show how quantum walks can be used to probe the universality of the eigenstate thermalisation hypothesis (ETH) relation. We find that whilst in C60 with a single walker, the ETH relation does not hold for node position projectors, it does hold for the average position, enforced by a symmetry of the Hamiltonian. The findings suggest a unified study of quantum walks and quantum self-thermalizations is natural and feasible.  
 \end{abstract}

\maketitle


\section{{\em G\MakeLowercase{eneral introduction}}}
Quantum walks are analogues of classical random walks, where instead of taking the time evolution to be a matrix of transition probabilities between nodes on a graph, the time evolution is given by a unitary matrix with complex transition amplitudes (see e.g.\ reviews~\cite{Aharonov_et_al, Venegas_Andraca_2012}). 
It has been argued that quantum walks are faster than their analogous classical walks in spreading around a graph, and could therefore compute certain properties of {\em a priori} unknown graphs faster (see e.g.\ Refs.~\cite{Kadian, Ambainis_element_distinctness, Childs_Subsetfinding, Magniez05quantumalgorithms, dhamapurkar2023quantum}). Moreover, Grover's quantum search algorithm, which searches special elements via evolving a quantum superposition over elements, can be implemented as a quantum walk, by adding self-loops to the special nodes~\cite{Edfarhi}. There are several recent experimental implementations of quantum walks on small quantum  systems, see e.g.\ Refs.~\cite{Acasiete_2020, RenatoP}.

Classical walks are frequently used to model thermalizations, such that their stationary distribution over nodes is the Gibbs state (also called Boltzmann or Canonical state) for which the probability of node $i$ is determined by its energy $E_i$ and the temperature $T$ via $p(i)=\exp(-\beta E_i)/\sum_i \exp(-\beta E_i)$, where $\beta=\frac{1}{k_B T}$. Such thermalizing random walks have proven useful for computation in the widely employed simulated annealing algorithm, wherein the computation is formulated as the task of minimising the energy $E_i$ and the $T$ associated with the thermalization is gradually lowered towards $T=0$, increasing the likelihood of low energy states~\cite{Kirkpatrick1983, Ingber1993}. 

\begin{figure}[h!]
\includegraphics[trim = 120 260 130 245, width=\linewidth, right]{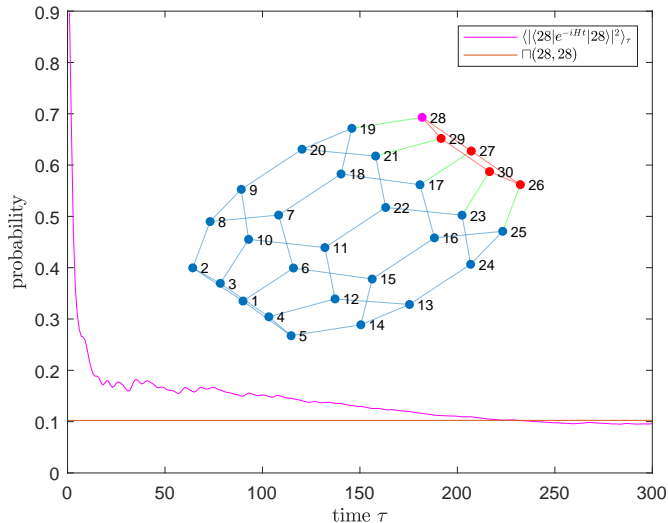}
\caption{\label{fig:full30} 
{{\bf Quantum equilibration of node probability.}} As an illustrative example, the time-averaged probability of a walk starting in the 28-th (\textcolor{magenta}{magenta}) node ending in the same node, $\langle |\bra{28}e^{-iHt}\ket{28}|^2\rangle_{\tau}$, equilibrates towards a limiting value $\mathcal{u}(28,28)$. The node in question  is part of a subsystem, a pentagon (\textcolor{red}{red}), and the remaining (mainly hexagonal) vertices are treated as the bath (\textcolor{blue}{blue}). The subsystem and bath are connected via interaction edges (\textcolor{green}{green}). The graph, with 30 vertices and 45 edges, is called fullerene 30. We analyse such equilibrations on graphs in general and on a wide family of fullerene graphs in more depth,  treating these equilibrations as self-thermalisations of unitarily evolving systems. }
\end{figure}

The usefulness of thermalization in classical walk computation together with the importance of quantum walk computation motivates investigating the connection between quantum walk computation and quantum thermalization. This is therefore the broad aim of this paper. More specifically, we ask: can quantum walks constitute quantum thermalization models and, if so, can we infer interesting properties of quantum walks using that knowledge?

We find that quantum walks do directly match models for quantum thermalization, more specifically, models of quantum thermalization of isolated quantum systems. A key reason for this connection is that the time-averaged state of the system is a central object both in quantum walks and thermalizations. We write this connection out and employ it to analyze the properties of the graph node position probability distribution of a quantum walk. 

For concreteness, we consider the family of graphs known as fullerenes (from fullerenes with 30 to 130 nodes), depicted in Fig.~\ref{fig:full30}. We show that a bound from Short et al~\cite{Short_2012} applies tightly to the equilibration of the graph node position observables. Nevertheless, we find that subsystems of the graph do not attain the Gibbs state $\gamma:=\frac{e^{-\beta H_S}}{\mathrm{tr}(e^{-\beta H_S})}$ as the time-averaged state.  As part of the Gibbs state analysis we explicitly describe how to define a sub-system Hamiltonian $H_s$ given the total graph Hamiltonian. We investigate whether observables related to the node position statistics obey the so-called eigenstate thermalization hypothesis~\cite{DAlessio_2015qtq, Deutsch_2018} relation, that the observable has (approximately) a particular neat form in the energy eigenbasis. Whilst the node projectors do not obey this relation for a single walker, the average node position observable surprisingly does, enabled by a symmetry of the Hamiltonian. We explain why these findings are all consistent. The results thus show quantum walks do, in a specific sense, correspond to quantum thermalization models and that the probability distribution over nodes can accordingly be analysed using tools designed for the study of quantum self-thermalization.  

We proceed as follows. We first describe quantum walks. We show how these are equivalent to quantum self-thermalization models. We then focus on the case of fullerene graphs. We show how to use quantum self-thermalisation tools to understand quantum walk equilibration. We define the Gibbs state and show it is not attained through the equilibration process. We investigate the ETH relation for observables related to the node statistics and finally give a discussion. 

\section{{\em Q\MakeLowercase{uantum walks}}}
We now describe the model for quantum walks we shall consider, following standard definitions such as those in Ref.~\cite{Farhi_1998}. We first define a  quantum walk on a graph followed by the so-called limiting distribution for the node position probabilities. 

A {\em continuous time quantum walk} on graph $G$ with ($N$) nodes labelled by $a=1,2,...N$ occurs in a Hilbert space of dimension $N$. The relation to {\em discrete time} quantum walks is described in Ref.~\cite{Childs_2010}. The basis of the Hilbert space is a set $\{\ket{a} \}$, with one orthogonal basis state $\ket{a}$ for each vertex $a$ in $G$, Thus, a general pure state $\ket{\psi(t)} = \sum_{a} \alpha_{a}(t) \ket{a}=e^{-iHt} \ket{\psi(0)}$, where $\alpha_{a}(t)$ is the probability amplitude to be at node $a$ at time $t$. The Hamiltonian $H$ is often taken to be the adjacency matrix $A$ defined by 
\begin{equation}\label{Eq:adjacency}
    \bra{a}A\ket{b} = \begin{cases}
       1 & a \neq b,  \text{if edge }(a,b) \in G \\
       0     & \text{otherwise}.
    \end{cases}
\end{equation}
For regular graphs, where each node has the same $d$ edges connected to it, one often chooses $H=\frac{A}{d}$. In either case, one sees by inspection that $H =H^T= H^\dagger$ where $T$ is the transpose operation and $\dagger$ is the hermitian conjugate operation. 

Measuring the node position at a random time $t\in [0,\tau]$ induces a probability distribution for transferring from the initial node $x$ to the final node $y$. More specifically, suppose we start in node $x$, run the continuous time quantum walk $U = e^{-iHt}$ for time $t$ chosen uniformly at random from $[0,\tau]$ and measure the node position. The probability of the measurement giving node $y$ is then given by the stochastic matrix $\Bar{P}$ with entries $\Bar{P}(x,y) = \frac{1}{\tau} \int_{0}^{\tau} |\bra{y} e^{-iHt}\ket{x}|^2 dt$.  Opening the integral shows $\Bar{P}(x,y)$ can be split into a time-dependent and time-independent part:
\begin{align}
&\Bar{P}(x,y)      = \frac{1}{N^2}\sum_{j} \Big|\sum_{k\in C_j}\braket{y}{\lambda_k}\braket{\lambda_k}{x} \Big|^2 +\\
 & \sum_{k,l: \lambda_k \neq \lambda_l} \braket{y}{\lambda_k}\braket{\lambda_l}{y}\braket{\lambda_k}{x}\braket{x}{\lambda_l} \frac{1}{\tau} \int_{0}^{\tau} e^{i(\lambda_l - \lambda_k)t} dt,
\end{align}
where $[\lambda_k, \ket{\lambda_k}]$ is the eigensystem of $A$ and $\{C_j\}$ is the partition of these indices $k$ obtained by adding all $k's$ that belong to the same $\lambda_k$. By inspection, as the maximal measurement time $\tau \rightarrow \infty $  the latter term in the above equation goes to zero,  resulting in what is termed the {\em limiting distribution} $\mathcal{u}$ with entries
\begin{equation}\label{Limiting}
     \mathcal{u}(x,y)=\frac{1}{N^2}\sum_{j} \Big|\sum_{k\in C_j}\braket{y}{\lambda_k}\braket{\lambda_k}{x} \Big|^2.
\end{equation}

We now move towards relating that limiting distribution with the time-averaged density matrix appearing in the literature on quantum thermalizations.

\section{{\em Q\MakeLowercase{uantum walks as self-thermalizations}}}
We first give an overview of relevant self-thermalization derivations. We then show how the position statistics in quantum walks on graphs, as defined above, fall within the domain of validity of specific results in the literature concerning thermal equilibration of observables of unitarily evolving quantum systems.  

A quantum system is commonly defined as being thermalised when its density matrix  approximately equals the Gibbs (also called Boltzmann or canonical) state for the given system Hamiltonian $H_S$: 
\begin{equation}\label{Gibbs}
 \gamma = \frac{e^{-\beta H_S}}{\mathrm{tr}(e^{-\beta H_S})}, 
\end{equation}
where $\beta = \frac{1}{k_B T}$, $k_B$ is the Boltzmann constant and $T$ is the temperature. (See the supplementary material C for a description of the relation between the Gibbs state and the so-called microcanonical ensemble). Equilibrium thermodynamics often assumes the system is in $\gamma$, and non-equilibrium thermodynamics that the system's state approaches $\gamma$ when interacting with a heat bath. A long-running question in the foundations of thermodynamics is how to justify these assumptions, assuming they are indeed justifiable (see e.g.\ Refs.~\cite{Gogolin_2016,Trushechkin_2022} for a review). In particular, in the case of an isolated quantum system, of which $S$ would be a subsystem, can one reasonably expect equilibration towards $\rho_S=\gamma$ given that the total system energy probabilities must be invariant in this case? 

Weaker notions of thermalisation may also be considered, such as certain observables having the same statistics as if the state were $\gamma$, or at least equilibrate to, or close to, some fixed value over time. In fact, several notions of thermalisation of observables are not tied to the Gibbs state, the equilibration towards which is indeed a strong assumption, but to more elementary notions of equilibration (see e.g.\ reviews \cite{Gogolin_2016, QuantumET}). We shall thus also investigate whether quantum walks exhibit such notions of thermalisation.

Many models of thermalisation, whether focused on the Gibbs state or not, can be viewed as telling a story about how information about the preparation of the system becomes inaccessible. In the case of typical entanglement arguments, one notes that typically quantum states are maximally entangled, meaning that any information about the preparation of the system is in global observables, i.e.\ observables that are not local on the system in question~\cite{schrodinger1989statistical, gemmer2003distribution, Goldstein_2006,popescu2006entanglement, Linden_2009, Dahlsten_2014}, an argument that can actually also be made in the classical case~\cite{muller2012unifying}. 

Another route which does not necessitate dividing the total system into subsystem and bath is to note that there are very many possible measurement bases and that if a system is prepared in one basis, by the uncertainty principle and the size of the Hilbert space, there will be many other bases in which the statistics look fully random, containing no information about the preparation. If we pick a random basis the statistics will thus likely look fully random regardless of the preparation. In this direction, there are recent successful derivations of a limited form of dynamical equilibration, showing that instantaneous expectation values of many observables $O$, $\mathrm{tr}(\rho(t) O)$, tend to be close to their time-averaged values. The time-averaged values can be written as~\cite{Short_2012,Linden_2009,Matteo,Reimann}
\begin{equation}
\label{eq:Otau}
\langle \mathrm{tr}(\rho(t) O) \rangle_{\tau \rightarrow \infty} := \lim_{\tau \rightarrow \infty}\frac{1}{\tau} \int_{0}^{\tau} \mathrm{tr}(\rho(t) O) dt = \mathrm{tr}(\omega O),   
\end{equation}
 where the time-averaged state

\begin{equation}
\label{eq:omega}
\omega =\langle \rho(t)\rangle_{\tau\rightarrow \infty}= \lim_{\tau \rightarrow \infty} \frac{1}{\tau} \int_{0}^{\tau} \rho(t) dt. 
\end{equation}

The time-averaged state $\omega$ is naturally associated with the idea  of equilibration of the system (see e.g.\ Ref.~\cite{reimann2010canonical} for a similar argument to what follows). The essential feature of equilibration is that expectation values of observables $O$ equilibrate to, or around,  some value that is roughly constant over time: $O_{eq}$. Then $O_{eq}= \langle \mathrm{tr}(\rho(t) O) \rangle_{\tau \rightarrow \infty}$, since if values equilibrate over time to constant values, those constant values must be the time averages of the values in question. Then, by Eq.~\ref{eq:Otau}, if there is equilibration of the system statistics, it is to the statistics of the time-averaged state $\omega$. (Separate assumptions and arguments, such as arguments given below, are needed to show that $\mathrm{tr}(\rho(t) O)$ is indeed close to  $O_{eq}$ at all times.) 

Now, to create a bridge between quantum walks and equilibration, we note a connection between the time-averaged state $\omega$ of Eq.~\ref{eq:omega} and the limiting distribution $\mathcal{u}$ of Eq.~\ref{Limiting} via the node position observable. Consider the set of node projectors
\[ \{O_x = \ketbra{x}{x} \vert x \in G \},\]
where $G$ is a graph with $N$ vertices. Let $\rho_0 =\ket{x}\bra{x}$ be the initial state. Then the expectation value of any observable $O_y = \ketbra{y}{y}$ with respect to the time-averaged state $\omega$ respects 
\begin{align}\label{equi_to_lim}
    \mathrm{tr}(O_y\omega) & = \lim_{\tau \rightarrow \infty} \frac{1}{\tau}\int_{0}^{\tau} \bra{y}e^{iAt}\ket{x}\bra{x}e^{-iAt}\ket{y}dt\\
     & = \frac{1}{N^2}\sum_{j} \Big|\sum_{k\in C_j}\braket{y}{\lambda_k}\braket{\lambda_k}{x}\Big|^2 \\
     & = \mathcal{u}(x,y).
\end{align}
Thus the probability of a walker being found in node $y$ after starting in node $x$, $\mathcal{u}(x,y)$, can be interpreted as the possible equilibrium value for the node observable $O_y$ of a system self-equilibrating under a unitary evolution. In fact, given the above connection, we can apply results from quantum thermalization theory to show that $\mathcal{u}(x,y)$ is an equilibrium value in the sense that the probability of being found at $y$, having started at $x$, is close to $\mathcal{u}(x,y)$ at most times.

In particular, Ref.~\cite{Short_2012} derived a  very general result about the concentration of the observable expectation values around the time average. They showed that for a system evolving under a general finite-dimensional Hamiltonian, in our case an $N$-dimensional quantum system associated with a graph $G$ with $N$ vertices, and any Hermitian observable $O$,  
\begin{equation}\label{Avg:bound}
    \langle |\text{tr}(O\rho(t)) - \text{tr}(O \omega)|^2\rangle_\tau \leq \frac{\parallel O\parallel^2 N(\epsilon)}{ d_{\text{eff}}} \left(1 + \frac{8 \log_{2}{N_\lambda}}{\epsilon \tau} \right).
\end{equation}
Starting from the left, $\langle .\rangle_\tau$ is the average over the interval $[0, \tau]$ and the time-averaged state $\omega$ is defined in Eq.\ref{eq:omega}. The operator norm $\parallel O\parallel^2$ is the largest singular value of $O$. $N(\epsilon)$, which concerns the density of energy gaps, is how many gaps between energy levels there are, within an energy interval of size $\epsilon$, maximised over energy intervals (see Ref.~\cite{Short_2012} for the precise definition). The effective dimension \begin{equation}\label{deff}
    d_{\text{eff}} \equiv \frac{1}{\sum_{n}(\mathrm{tr}(P_n\rho(0)))^2}
\end{equation} tells us the inverse purity ($\mathrm{tr}(\rho^2)$) of $\rho$ in the energy basis, with $P_n$ being the projector onto the $n$-th energy eigenspace~\cite{Short_2012}. Finally, $N_{\lambda}$ is the number of distinct energy eigenvalues. Given the connection to quantum walks via Eq.~\ref{equi_to_lim}, the bound of Eq.~\ref{Avg:bound} can be applied to observables measured at the end of the quantum walk, in particular the node projectors.

\section{{\em Q\MakeLowercase{uantum walks on fullerene graphs}}}
For concreteness, we will apply the general connection between quantum walks and thermalisation identified above to the illustrative special case of {\em quantum walks on fullerene graphs}. Choosing a particular case allows us to undertake numerical experiments as part of the analysis. 

Fullerene graphs can be thought of as a generalisation of the famous buckyball C60 graph. (It is safest not to interpret these graphs as models of physical fullerene molecules which are much more complicated.). The buckyball graph is important from a  mathematical point of view because of its symmetries. It is a Cayley graph generated by a particular symmetric subset of the alternating group of five elements~\cite{bucky1}. 

Fullerenes have fixed numbers of pentagons and varying numbers of hexagons. There are 12 pentagons whereas the number of hexagons can take all values from $\mathrm{N}$ except 1. Fullerene graphs satisfy the Euler formula $  f - e + v = 2$ for polyhedrons with $f$ faces, $e$ edges, and $v$ vertices~\cite{gm}. Fullerene graphs with $N$ vertices exist for any even $N \geq 20$ except $N = 22$. 

We used the database from Ref.~\cite{BRINKMANN} of fullerene graphs with vertices 30 to 130 in multiples of ten (except 22). We define the graph as a tuple, for example, $F_{20} = F(12,0)$, where the first value is the number of pentagons and the second is the number of hexagons. In general $F_{10(k+1)} = F(12,5(k-1))$ for $k\geq 1$. The number of hexagons and pentagons does not uniquely specify the fullerene graph-there are isomers. We shall perform several simulations on a well-known  isomer of $F_{60}$, the C60 (buckyball) graph (depicted in Fig.~\ref{fig:avg-bound}). To consider increasing bath sizes, we shall also simulate isomers of $F_{30}$, $F_{40}$, \dots, $F_{130}$ that form a closed tube-like shape. 

\subsection{{\em A\MakeLowercase{nalytical understanding of how position  probabilities equilibrate}}}
\begin{figure}[htbp]
    \centering
    \includegraphics[trim=50 230 110 235,clip,width=0.55\textwidth,keepaspectratio, right]{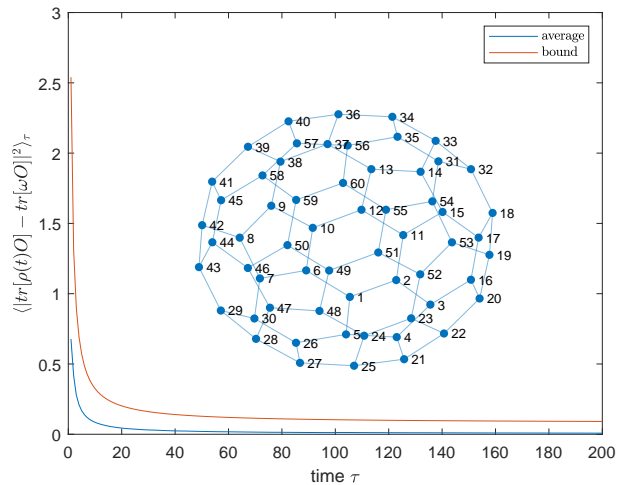} \caption{{{\bf Quantum equilibration bound is \textcolor{blue}{tight}.}} The bound given in Eq.~\ref{Avg:bound} holds for the buckyball (fullerene 60) case where $O=\ket{1}\bra{1}$ is the projector onto node 1. One sees that the expectation value of the projector becomes very close to its time-average equilibrium value and that the analytical bound captures this behaviour very well. }
    \label{fig:avg-bound}
\end{figure}

Adapting the bound from Ref.\ \cite{Short_2012}, given in Eq.~\ref{Avg:bound} to the buckyball gives 
\begin{equation}
\langle |\text{tr}(O\rho(t)) - \text{tr}(O\omega)|^2\rangle_{\tau} \leq 0.08\left(1+ \frac{8\times3.90}{\tau}\right).
\end{equation}
Here $\log_{2}{N_\lambda} = 3.90$, $d_{\text{eff}} = 1/0.08$ (we write the initial state $\ket{1}$ in the energy eigenbasis of $H$ and then calculate $d_{\text{eff}}$ numerically using Eq. \ref{deff}.), $||O||^2 = 1$, and $N(\epsilon) = 1$ for $\epsilon =1 $. Fig.~\ref{fig:avg-bound} shows that the resulting bound, applied to a node position projector, $\ket{1}\bra{1}$, is surprisingly tight. This implies that the probability of being in node $y$ given that one started in node $x$ equilibrates to the limiting distribution value $\mathcal{u}(x,y)$ in the sense of being near that value at most times for large times $\tau$.
\subsection{{\em T\MakeLowercase{he limiting distribution has symmetries and initial state dependence}}} 
\begin{figure}[htbp]
        \centering
        \includegraphics[trim = 100 250 50 235,width=0.55\textwidth,keepaspectratio, left]{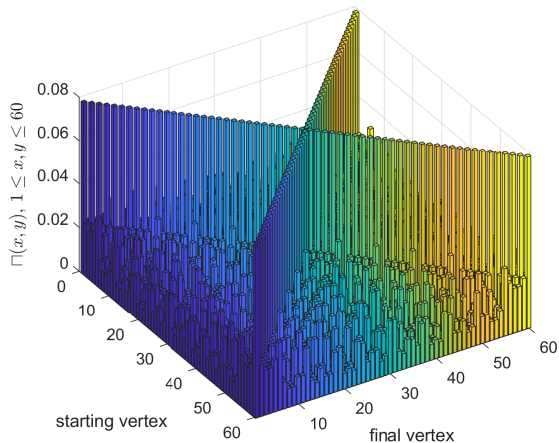}
        \caption{{{\bf Position distribution ( time averaged over $\tau\rightarrow \infty$) for C60 graph}} Each vertex is treated as a starting point. The z-axis shows the limiting distribution $\mathcal{u}(x,y)$ values when we start the quantum walk at vertex $x$ and measure at $y$ for all $1 \leq x, y \leq 60$. The graph shows highly nonuniform behaviour as well as some symmetries. This illustrates strong initial-state dependence of the distribution of the final position.}
        \label{fig:buckyball3d}
\end{figure}

We now show and analyse the position distribution over final positions: the limiting distribution (Eq.~\ref{Limiting}). The limiting position distribution for a quantum walk on the C60 graph, shown in Fig.~\ref{fig:buckyball3d}, is highly uneven and has certain notable features. 

A notable feature is an X-like shape. Analytically, we observe that for any node $x$, $1 \leq x \leq 60$, 
\begin{equation}\label{equal}
    \braket{x}{\lambda_k} = \pm \braket{61-x}{\lambda_k}
\end{equation}
for all eigenstates of $A$ (Refer to the supplementary material Appendix D for the proof). Eq.~\ref{Limiting} can be used to explain this X. The limiting distribution corresponding to a particular starting node $x$ has two peaks at $\mathcal{u}(x,x)$ and $\mathcal{u}(x,61-x)$ and both are equal by Eq.~\ref{equal}. This gives us symmetry around node 30. Hence, for every $1 \leq y \leq 60$ we have $\mathcal{u}(x,y) = \mathcal{u}(x,61-y)$. This gives rise to the cross-like pattern in Fig.~\ref{fig:buckyball3d}.
The limiting distribution of Fig.~\ref{fig:buckyball3d}, by inspection, exhibits strong initial state dependence, in that $\mathcal{u}(x,y)$ depends heavily on $x$. Thus the time-averaged probability of $y$ depends heavily on the starting node $x$. This initial state dependence poses a challenge to the idea that a subsystem may be reset to the Gibbs state by this process.

\section{{\em G\MakeLowercase{ibbs state on subsystem (pentagon)}}}    
In anticipation of comparing the time-averaged state on a subsystem (a pentagon) with the Gibbs state, we derive the Gibbs state (Eq.~\ref{Gibbs}) for such a subsystem.

Consider firstly how to define the system Hamiltonian $H_s$ in the case of a quantum walk on a fullerene graph. There are many possible choices of subsystems and we choose that of a pentagon since the number of pentagons is fixed whereas the number of hexagons scales up with the fullerene number, allowing for a natural scaling of the bath size. 

\begin{figure}[h]
    \centering
    \begin{tikzpicture}
\draw (0,0) -- (3,0);
\filldraw [black] (0,0) circle (1pt)node[anchor=south]{$\ket{1}_1\ket{0}_{2}$};
\filldraw [black] (0,0) circle (0pt)node[anchor=north]{$1$};
\filldraw [black] (3,0) circle (1pt)node[anchor=south]{$\ket{0}_1\ket{1}_2$};
\filldraw [black] (3,0) circle (0pt)node[anchor=north]{$2$};
\end{tikzpicture} 
    \caption{{{\bf Single excitation example.}} The walker being at a given node is associated with a basis state of a product basis with each node as a sub-system.}
    \label{fig:toy_example}
\end{figure}
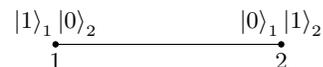

To define the basis states of the system and subsystem we find it helpful to think of the walk physically as a single excitation (the walker) moving around the graph. For some reason, states with several walkers are effectively banned. To illustrate how we split the graph into subsystems, consider, as a warm-up, a graph with two vertices $1$ and $2$ sharing one edge as a total system and a single walker (depicted in Fig. \ref{fig:toy_example}). Each vertex is a sub-system with two states $\ket{0}$ and $\ket{1}$ corresponding to no walker on the node and one walker on the node respectively. The product basis is thus $\mathcal{B} = \{ \ket{0}_1\ket{0}_2, \ket{0}_1\ket{1}_2, \ket{1}_1\ket{0}_2, \ket{1}_1\ket{1}_2\}$. Under the single-walker restriction, the state is always a superposition of $\ket{0}_1\ket{1}_2$ and  $\ket{1}_1\ket{0}_2$ and the Hamiltonian has no terms coupling to the other two states. 

Generalising to any graph, the single walker state space is spanned by  $\mathcal{B} = \{ \ket{b}_j = \ket{0}_1\ket{0}_2\dots\ket{1}_{j}\dots \ket{0}_N \vert 1\leq j \leq N \}$. On the pentagon $S$ or the bath $B$ (the remaining nodes), there is also the possibility of having no walker, such that we should also include extra basis elements $\ket{b}_0^S=\ket{0}_1\ket{0}_2\ket{0}_3\ket{0}_4\ket{0}_5$ and $\ket{b}_0^B=\ket{0}_6\ket{0}_7\dots \ket{0}_N$ in their respective basis. 

We break up the total Hamiltonian into system, bath and interaction Hamiltonians: $\Tilde{H}_{tot}= \Tilde{H}_S + \Tilde{H}_B + \Tilde{H}_{int}$, where $\Tilde{}$ indicates  acting on the whole graph.

The total Hamiltonian $\Tilde{H}_{tot}$ is given by the adjacency matrix of the graph. Each vertex of the fullerene graph $F_{N}$ has degree three and
\begin{equation}\label{eq:totH}
    \Tilde{H}_{tot}\ket{b}_j = (\ket{b}_{r}+\ket{b}_{s} + \ket{b}_{t}), 
\end{equation}
where $1 \leq r,s,t \leq N$ and $r$,$s$,$t$ are neighbours of $j$. 

$H_S$ encodes the connection between vertices on the pentagon of interest. $H_B$ encodes the connection between vertices on hexagons and remaining pentagons within the bath. $H_{int}$ encodes the connection between vertices in the pentagon of interest and {\em the remaining} vertices. For any $N$, $H_S$ and $H_{int}$ are invariant. Only $H_B$ changes as we consider different $N$'s. 

The Hamiltonian on the pentagon subsystem,  $\Tilde{H}_S := H_S \otimes I_B$, is the part of the total Hamiltonian acting trivially on the bath. $H_S$ is given by
\begin{equation}\label{subsystem}
    \bra{b}^S_k H_S\ket{b}^S_j = \begin{cases}
       1 &  \ket{b}^S_j \neq \ket{b}^S_k, \mathrm{edge}(j,k) \in F_N\\
        &  j,k \in \{1,2,3,4,5\}\\
       0     & \text{otherwise},
    \end{cases}
\end{equation}
where $\mathcal{H}_S = \mathrm{span}\{\ket{b}^S_j \vert 1 \leq j \leq 5 \}$. 

\begin{equation}\label{eq:nowalkersys}
  \begin{aligned}
   \bra{b}^S_0 H_S\ket{b}^S_0  & = 0\\
   \bra{b}^S_0 H_S\ket{b}^S_j  & = 0\\
   \bra{b}^S_k H_S\ket{b}^S_0  & = 0
\end{aligned}  
\end{equation}
for $j,k \in \{1,2,3,4,5\}$. For completeness, we also define $\Tilde H_{B}$ and $\Tilde H_{int}$ in the supplementary material A. 

We define the Gibbs state within the single-excitation subspace, in line with the quantum walk model. We call the space $\mathcal{H}_S$ spanned by the subsystem (pentagon) basis $\{\ket{b}^S_j \vert 1 \leq j \leq 5 \} \cup \ket{b}^S_0 $ the subsystem space. 


Having defined $H_S$ in Eq.~\ref{subsystem} the Gibbs state on the pentagon immediately follows. It is a $6\times6$ matrix 
and the eigenvalues and eigenvectors can be derived in a two-stage process: (i) note that $\ket{b}^S_0$ is an eigenvector with eigenvalue 0 and that $H_S$ is, in the node basis including $\ket{b}^S_0$, a block diagonal composition
\[H_S =(0\ket{b}^S_0\bra{b}^S_0) \oplus A_5=\begin{bmatrix} 0 & 0 \\
   0 & A_5\end{bmatrix},\]
where $A_{5}(j,k) =\bra{b}^S_k H_S \ket{b}^S_j/2$ is the normalised adjacency matrix of the pentagon, (ii) calculate the additional eigenvectors and eigenvalues for $A_{5}$. The eigenvalues of $H_S$ are $0$ and $\lambda_j = \cos{(2 \pi j/5)}$ for $ j \in \{0,1,2,3,4\}$. The corresponding eigenvectors are $\ket{b}^S_0 = [1\, 0\, 0\, 0\, 0\, 0 ]^{\dagger}$ and $\ket{\lambda_j} = (1/\sqrt{5})[0, 1, \omega^j, \omega^{2j},\omega^{3j},\omega^{4j}],$  where $\omega = e^{\frac{2 \pi i}{5}}$. Inserting $H_S=\sum_j \lambda_j\ketbra{\lambda_j}{\lambda_j}$ into the definition of a Gibbs state (Eq.~\ref{Gibbs}) yields
\begin{equation}
    \begin{aligned}\label{eq:gammas}
    \gamma_S & = 
    \frac{1}{Z} \ketbra{b^S_0}{b^S_0} + \frac{1}{Z}\sum_{j = 0}^{4}\exp{(-\beta \cos{(2 \pi j/5)})} \ketbra{\lambda_j}{\lambda_j}.
\end{aligned}
\end{equation}
The partition function
\begin{equation}
\label{eq:Z}
Z = e^{-\beta} + 2 e^{((1 + \sqrt{5})\beta/4)} + 2 e^{(-(\sqrt{5}-1)\beta/4)} + 1.
\end{equation}

Equations \ref{eq:gammas} and \ref{eq:Z} give a clear understanding of the Gibbs state and in particular the node probabilities associated with it. For any temperature, $\gamma_S$ assigns the same probability for all nodes $j$. The probability of being at node $j$ in the node basis $\{\ket{j} \vert j \in \{1,2,3,4,5\}\}$, 
\begin{equation}
  \begin{aligned}\label{eq:pj}
    p(j) &=\bra{j} \gamma_S \ket{j}\\
   & =
\bra{j}\left( \frac{1}{Z}\sum_{j = 0}^{4}\exp{(-\beta \cos{(2 \pi j/5)})} \ketbra{\lambda_j}{\lambda_j}\right)\ket{j}\\
&=\frac{1}{Z}\sum_{j = 0}^{4}\frac{\exp{(-\beta \cos{(2 \pi j/5)})}}{5},
\end{aligned}  
\end{equation}
which is independent of $j$. For $\beta \gg 1$, $p(j)\approx 0.2$  for $j \in \{1,2,3,4,5\}$ and the probability of no walker in the pentagon, $\bra{b}^S_0 \gamma_S \ket{b}^S_0 \approx 0$. For $\beta \ll 1$, $p(j)\approx \frac{1}{Z}\approx 1/6$ for $j \in \{1,2,3,4,5\}$. Moreover, $\bra{b}^S_0 \gamma_S \ket{b}^S_0 \approx 1/6$. As expected, at very large temperatures, the Gibbs state effectively becomes the maximally mixed state. (the details for the extreme cases of $\beta$ values are given in the supplementary material A).

\subsection{{\em C\MakeLowercase{omparison of \MakeUppercase{G}ibbs state and time-averaged state on the pentagonal subsystem}}}    
To disprove that the subsystem time-average state is the Gibbs state, it is sufficient to show that for some sets of observables, the statistics are inconsistent with those of Gibbs states. We focus on the probabilities of being at the nodes. We first consider the probability of being at the $N$-th node, followed by an analysis of whether the node probability distribution of the time-averaged state, within the pentagonal subsystem, has initial state dependence.

\begin{figure}[htbp]
        \centering
\includegraphics[trim = 100 250 80 235, width=0.52\textwidth,keepaspectratio, left]{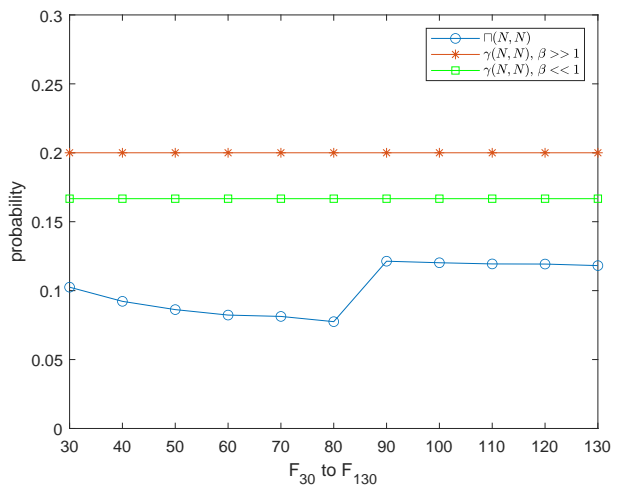}
        \caption{{{\bf Probability of ending in node $N$(pentagonal node).}} The time-averaged probability $\mathcal{u}(N,N)$ and possible Gibbs state probabilities, $\gamma_{N,N}$, from  Eqs.\ref{eq:Z} and \ref{eq:pj} are plotted for $F_{30}$ to $F_{130}$. One sees that they are not equal for any $N$. }
        \label{fig:subsystemplot}
    
    \end{figure}

Consider the probability of the final measurement finding the walker at the $N$-th node. The $N$-th node lies within the pentagonal subsystem for fullerene graphs $F_N$ for any $N$. We run the quantum walk of $\Tilde{H}$ given in Eq.~\ref{eq:totH}, for very large $\tau$, starting in node $\ket{N}$. We evaluate $\mathcal{u}(N,N)$, the limiting distribution entry giving the probability of being in node $N$ at the end. We compare this to the probability of being in node $N$ assigned by the Gibbs state $\gamma$, which we derived in Eq.~\ref{eq:pj}.  The numerical experiment includes $F_{30}$ to $F_{130}$ in multiples of ten, corresponding to scaling the bath size. The resulting plot in Fig.\ref{fig:subsystemplot} shows a clear inconsistency between the Gibbs state and the time-averaged statistics even as the bath size is increased. 

Another method of seeing that the time-equilibrium state on the subsystem does not equal the Gibbs state is to make use of the initial state dependence. 
For example, if $x=1$, and $x=2$ respectively the limiting distribution vectors, with entries associated with the different y's within the pentagon, are
\begin{eqnarray*}
   &[0.079, 0.024, 0.021, 0.021, 0.024]& \textrm{and} \\
   &[0.024, 0.079, 0.024, 0.021, 0.021]&,   
\end{eqnarray*}
 (up to 3 decimal places). In contrast, the $y$ probabilities from the Gibbs state are independent of the initial state. 
 
We conclude that the Gibbs state on the Pentagon subsystem does not equal the time-averaged state, even as the number of nodes $N$ is scaled up. Moreover, the subsystem's node position statistics differ from those of the Gibbs state.

\section{{\em E\MakeLowercase{igenstate thermalisation hypothesis and node positions}}}    
In this section, we first describe the eigenstate thermalisation hypothesis (ETH). We then show that the ETH is, for a case of a single walker, violated for the projectors onto individual nodes but respected for the average position.  

The eigenstate thermalisation hypothesis can be stated as a relation that a system Hamiltonian $H$ and some observable $O$ of interest may jointly satisfy. The relation essentially states that $O$ takes (at least approximately) the neat form of a diagonal matrix with smoothly changing values on the diagonal, when written in the Hamiltonian eigenbasis $\{\ket{\lambda_m} \}$ (with the eigenstates ordered such that $E_m\geq E_{m-1}$)~\cite{DAlessio_2015qtq, Deutsch_2018, Gogolin_2016,Dunlop_2021}. In other words, roughly speaking,  the relation is satisfied whenever $\bra{\lambda_m}O\ket{\lambda_n}\approx f(m)\delta_{mn}$ for all $m$, $n$, where $f(m)$ changes slowly in $m$. (For an example of a more precise version of the ETH relation see Ref.~\cite{DAlessio_2015qtq}). The ETH relation is similar to the relation that many observables have with
a random basis~\cite{DAlessio_2015qtq}.  

Under the ETH relation, a peculiar property of energy eigenstates can be extended to more general initial states. If an energy eigenstate $\ket{\lambda_{m'}}$ is the initial state, the time evolution under $H$ is trivial, adding a global phase. Then the instantaneous average $\langle O\rangle(t)=O_{m'm'}$ at any time $t$. 

Now generalize the starting state to be any pure state with support (non-zero amplitudes) only within a narrow window around some energy $E_{m'}$. For general states $\ket{\psi(t)}=\sum_m c_m(t=0)e^{-iE_mt}\ket{m}$ (taking $\hbar=1$) such that 
\begin{align*}
 &\langle O(t)\rangle =\bra{\psi(t)}O\ket{\psi(t)}=\\
  & \sum_m |c_m(t=0)|^2 O_{mm}\!+\!\sum_{m\neq n} c^*_mc_nO_{mn}e^{-i(E_m-E_n)t}.  
\end{align*}
Hence, if the ETH relation described above holds, then~\cite{Deutsch_2018}
\begin{equation}
\label{eq:equalsmc}
\langle O\rangle(t)\approx O_{m'm'}.
\end{equation} 

We now tackle the question of whether the ETH relation is satisfied by observables associated with the graph node positions. More specifically,  if we start localised in a node, $\ket{x}$, do the average position observable $O = \sum_{x} x \ket{x}\bra{x}$, or the projectors $O_x=\ket{x}\bra{x}$ satisfy the ETH relation? 

\begin{figure}[htbp]
        \centering
        \includegraphics[trim = 90 260 70 235,width=0.55\textwidth,keepaspectratio, left]{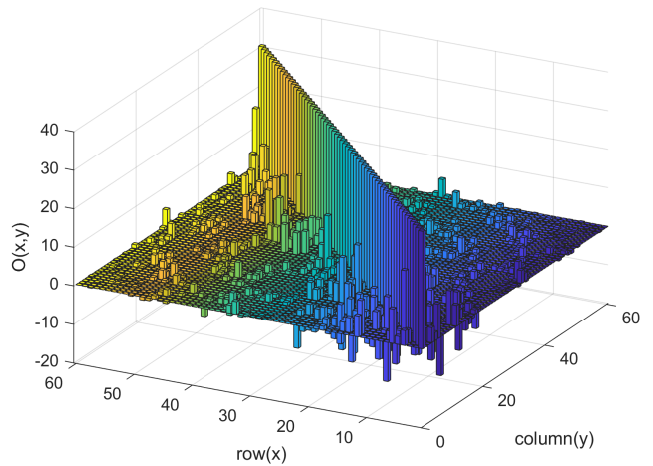}
        \caption{ {{\bf Position observable in Hamiltonian eigenbasis:}} We plot the matrix elements of $O = \sum_{x = 1}^{60} x\ketbra{x}{x}$ in the energy eigenbasis of $H$, for the C60 graph. Observe that the diagonal of $O$ in energy eigenbasis is a smooth function (almost constant) and there are only small off-diagonal elements, in line with the ETH relation. }
        \label{fig:ETHplot}    
    \end{figure}

We show that the ETH does not hold for the projectors $O_x=\ket{x}\bra{x}$ (see supplementary material B for details and analysis). We find that the ETH relation is nevertheless, perhaps surprisingly, qualitatively respected by the position observable  $O = \sum_{x = 1}^{60} x \ketbra{x}{x}$. The symmetry of the graph turns out to enhance the validity of the relation. $O$ represented in the energy eigenbasis is depicted in Fig.~\ref{fig:ETHplot}. 
The diagonal values can be analytically calculated from the symmetry of Eq.~\ref{equal}: 
\begin{align}
   O_{m,m} &= \sum_{x = 1}^{60} x \braket{\lambda_m}{x}\braket{x}{\lambda_m} \\
   & = \sum_{x = 1}^{30} (x+(61-x))|\braket{\lambda_m}{x}|^2\\
   & = 61 \sum_{x = 1}^{30} |\braket{\lambda_m}{x}|^2\\
   & = 61/2\\
   & = 30.5.
\end{align}
The symmetry of Eq.~\ref{equal} thus has the impact of removing the fluctuations around 30.5 that one would expect if the amplitudes of node states in the energy basis
$\braket{\lambda_m}{x}$ were i.i.d. (or approximately so) with respect to $x$. This phenomenon enhances the validity of the ETH relation for $O$. It should be investigated whether the ETH relation holds if there are many walkers. 


\section{{\em S\MakeLowercase{ummary and outlook}}}
We showed how to associate quantum walks over graphs with self-thermalizations. The resulting thermalization models turn out to fall within the domain of validity of several recent results concerning the self-thermalization of unitarily evolving quantum systems. We illustrate this in the case of fullerene graphs with single walkers. Some notions of thermalisation take place whilst some do not, as summarised in Table~I.
The results create a concrete bridge between the study of self-thermalization of quantum systems and that of quantum computation via quantum walks. Perhaps the most concrete benefit of this bridge so far is the finding that certain equilibration bounds can be applied to quantum walks.

\begin{table*}\label{table:summary}
  \centering
\begin{tabular}{|c|c|}\hline
 {\bf Thermalisation criterion} & {\bf Satisfied?} \\ \hline
 Expectation value equilibration bounds like Refs~\cite{reimann2010canonical, Short_2012} hold & Yes\\ \hline
 Equilibration to Gibb's state of a subsystem & No\\ \hline
Equilibration removing initial state dependence & No\\ \hline
Eigenstate thermalisation hypothesis
 & Yes for $\langle X \rangle$, no for $\langle \ket{x}\bra{x}\rangle$\\ \hline
\end{tabular}
  \caption{Which types of thermalisation were found to take place and which did not for single walker quantum walks.}\label{compare}
\end{table*}

We expect that there are many more ways to build on and make use of this bridge. For example it would be interesting to investigate the quantum walk Grover's algorithm~\cite{Edfarhi} (time-averaged) as a kind of thermalization,  and compare that with the physical interpretation of the algorithm as a type of ballistic descent into (and out of) a potential minimum~\cite{grover2001schrodinger}. This may yield better physical insight into the apparent speed-up of quantum search. Since there is a procedure to simulate continuous time quantum walks on graphs via discrete time coined walks~\cite{Childs_2010}, we moreover anticipate that the approach here can be extended to the latter type of quantum walks. Finally, these investigations, including the validity of the ETH hypothesis, should be extended to the case of many walkers. 




\noindent {\bf {\em Acknowledgements.---}} We gratefully acknowledge discussions with Meng Fei, Libo Jiang, Viv Kendon, Xiao Li and Zhang Zedong. This work was supported by the National Natural Science Foundation of China (Grants No. 12050410246, No. 1200509, No. 12050410245) and City University of Hong Kong (Project No. 9610623).

\bibliographystyle{apsrev4-1}
\bibliography{apssamp}

\providecommand{\noopsort}[1]{}\providecommand{\singleletter}[1]{#1}%
\begin{thebibliography}{39}%
\makeatletter
\providecommand \@ifxundefined [1]{%
 \@ifx{#1\undefined}
}%
\providecommand \@ifnum [1]{%
 \ifnum #1\expandafter \@firstoftwo
 \else \expandafter \@secondoftwo
 \fi
}%
\providecommand \@ifx [1]{%
 \ifx #1\expandafter \@firstoftwo
 \else \expandafter \@secondoftwo
 \fi
}%
\providecommand \natexlab [1]{#1}%
\providecommand \enquote  [1]{``#1''}%
\providecommand \bibnamefont  [1]{#1}%
\providecommand \bibfnamefont [1]{#1}%
\providecommand \citenamefont [1]{#1}%
\providecommand \href@noop [0]{\@secondoftwo}%
\providecommand \href [0]{\begingroup \@sanitize@url \@href}%
\providecommand \@href[1]{\@@startlink{#1}\@@href}%
\providecommand \@@href[1]{\endgroup#1\@@endlink}%
\providecommand \@sanitize@url [0]{\catcode `\\12\catcode `\$12\catcode
  `\&12\catcode `\#12\catcode `\^12\catcode `\_12\catcode `\%12\relax}%
\providecommand \@@startlink[1]{}%
\providecommand \@@endlink[0]{}%
\providecommand \url  [0]{\begingroup\@sanitize@url \@url }%
\providecommand \@url [1]{\endgroup\@href {#1}{\urlprefix }}%
\providecommand \urlprefix  [0]{URL }%
\providecommand \Eprint [0]{\href }%
\providecommand \doibase [0]{http://dx.doi.org/}%
\providecommand \selectlanguage [0]{\@gobble}%
\providecommand \bibinfo  [0]{\@secondoftwo}%
\providecommand \bibfield  [0]{\@secondoftwo}%
\providecommand \translation [1]{[#1]}%
\providecommand \BibitemOpen [0]{}%
\providecommand \bibitemStop [0]{}%
\providecommand \bibitemNoStop [0]{.\EOS\space}%
\providecommand \EOS [0]{\spacefactor3000\relax}%
\providecommand \BibitemShut  [1]{\csname bibitem#1\endcsname}%
\let\auto@bib@innerbib\@empty
\bibitem [{\citenamefont {Aharonov}\ \emph {et~al.}(2001)\citenamefont
  {Aharonov}, \citenamefont {Ambainis}, \citenamefont {Kempe},\ and\
  \citenamefont {Vazirani}}]{Aharonov_et_al}%
  \BibitemOpen
  \bibfield  {author} {\bibinfo {author} {\bibfnamefont {D.}~\bibnamefont
  {Aharonov}}, \bibinfo {author} {\bibfnamefont {A.}~\bibnamefont {Ambainis}},
  \bibinfo {author} {\bibfnamefont {J.}~\bibnamefont {Kempe}}, \ and\ \bibinfo
  {author} {\bibfnamefont {U.}~\bibnamefont {Vazirani}},\ }in\ \href@noop {}
  {\emph {\bibinfo {booktitle} {Proceedings of the thirty-third annual ACM
  symposium on Theory of computing}}}\ (\bibinfo {year} {2001})\ pp.\ \bibinfo
  {pages} {50--59}\BibitemShut {NoStop}%
\bibitem [{\citenamefont {Venegas-Andraca}(2012)}]{Venegas_Andraca_2012}%
  \BibitemOpen
  \bibfield  {author} {\bibinfo {author} {\bibfnamefont {S.~E.}\ \bibnamefont
  {Venegas-Andraca}},\ }\href@noop {} {\bibfield  {journal} {\bibinfo
  {journal} {Quantum Information Processing}\ }\textbf {\bibinfo {volume}
  {11}},\ \bibinfo {pages} {1015} (\bibinfo {year} {2012})}\BibitemShut
  {NoStop}%
\bibitem [{\citenamefont {Kadian}\ \emph {et~al.}(2021)\citenamefont {Kadian},
  \citenamefont {Garhwal},\ and\ \citenamefont {Kumar}}]{Kadian}%
  \BibitemOpen
  \bibfield  {author} {\bibinfo {author} {\bibfnamefont {K.}~\bibnamefont
  {Kadian}}, \bibinfo {author} {\bibfnamefont {S.}~\bibnamefont {Garhwal}}, \
  and\ \bibinfo {author} {\bibfnamefont {A.}~\bibnamefont {Kumar}},\
  }\href@noop {} {\bibfield  {journal} {\bibinfo  {journal} {Computer Science
  Review}\ }\textbf {\bibinfo {volume} {41}},\ \bibinfo {pages} {100419}
  (\bibinfo {year} {2021})}\BibitemShut {NoStop}%
\bibitem [{\citenamefont {Ambainis}(2007)}]{Ambainis_element_distinctness}%
  \BibitemOpen
  \bibfield  {author} {\bibinfo {author} {\bibfnamefont {A.}~\bibnamefont
  {Ambainis}},\ }\href@noop {} {\bibfield  {journal} {\bibinfo  {journal} {SIAM
  Journal on Computing}\ }\textbf {\bibinfo {volume} {37}},\ \bibinfo {pages}
  {210} (\bibinfo {year} {2007})}\BibitemShut {NoStop}%
\bibitem [{\citenamefont {Childs}\ and\ \citenamefont
  {Eisenberg}(2003)}]{Childs_Subsetfinding}%
  \BibitemOpen
  \bibfield  {author} {\bibinfo {author} {\bibfnamefont {A.~M.}\ \bibnamefont
  {Childs}}\ and\ \bibinfo {author} {\bibfnamefont {J.~M.}\ \bibnamefont
  {Eisenberg}},\ }\href@noop {} {\bibfield  {journal} {\bibinfo  {journal}
  {arXiv preprint quant-ph/0311038}\ } (\bibinfo {year} {2003})}\BibitemShut
  {NoStop}%
\bibitem [{\citenamefont {Magniez}\ \emph {et~al.}(2007)\citenamefont
  {Magniez}, \citenamefont {Santha},\ and\ \citenamefont
  {Szegedy}}]{Magniez05quantumalgorithms}%
  \BibitemOpen
  \bibfield  {author} {\bibinfo {author} {\bibfnamefont {F.}~\bibnamefont
  {Magniez}}, \bibinfo {author} {\bibfnamefont {M.}~\bibnamefont {Santha}}, \
  and\ \bibinfo {author} {\bibfnamefont {M.}~\bibnamefont {Szegedy}},\
  }\href@noop {} {\bibfield  {journal} {\bibinfo  {journal} {SIAM Journal on
  Computing}\ }\textbf {\bibinfo {volume} {37}},\ \bibinfo {pages} {413}
  (\bibinfo {year} {2007})}\BibitemShut {NoStop}%
\bibitem [{\citenamefont {Dhamapurkar}\ and\ \citenamefont
  {Deng}(2023)}]{dhamapurkar2023quantum}%
  \BibitemOpen
  \bibfield  {author} {\bibinfo {author} {\bibfnamefont {S.}~\bibnamefont
  {Dhamapurkar}}\ and\ \bibinfo {author} {\bibfnamefont {X.-H.}\ \bibnamefont
  {Deng}},\ }\href@noop {} {\bibfield  {journal} {\bibinfo  {journal} {Physica
  A: Statistical Mechanics and its Applications}\ }\textbf {\bibinfo {volume}
  {630}},\ \bibinfo {pages} {129252} (\bibinfo {year} {2023})}\BibitemShut
  {NoStop}%
\bibitem [{\citenamefont {Farhi}\ and\ \citenamefont
  {Gutmann}(1998{\natexlab{a}})}]{Edfarhi}%
  \BibitemOpen
  \bibfield  {author} {\bibinfo {author} {\bibfnamefont {E.}~\bibnamefont
  {Farhi}}\ and\ \bibinfo {author} {\bibfnamefont {S.}~\bibnamefont
  {Gutmann}},\ }\href@noop {} {\bibfield  {journal} {\bibinfo  {journal}
  {Physical Review A}\ }\textbf {\bibinfo {volume} {57}},\ \bibinfo {pages}
  {2403} (\bibinfo {year} {1998}{\natexlab{a}})}\BibitemShut {NoStop}%
\bibitem [{\citenamefont {Acasiete}\ \emph {et~al.}(2020)\citenamefont
  {Acasiete}, \citenamefont {Agostini}, \citenamefont {Moqadam},\ and\
  \citenamefont {Portugal}}]{Acasiete_2020}%
  \BibitemOpen
  \bibfield  {author} {\bibinfo {author} {\bibfnamefont {F.}~\bibnamefont
  {Acasiete}}, \bibinfo {author} {\bibfnamefont {F.~P.}\ \bibnamefont
  {Agostini}}, \bibinfo {author} {\bibfnamefont {J.~K.}\ \bibnamefont
  {Moqadam}}, \ and\ \bibinfo {author} {\bibfnamefont {R.}~\bibnamefont
  {Portugal}},\ }\href@noop {} {\bibfield  {journal} {\bibinfo  {journal}
  {Quantum Information Processing}\ }\textbf {\bibinfo {volume} {19}},\
  \bibinfo {pages} {1} (\bibinfo {year} {2020})}\BibitemShut {NoStop}%
\bibitem [{\citenamefont {Portugal}\ and\ \citenamefont
  {Moqadam}(2022)}]{RenatoP}%
  \BibitemOpen
  \bibfield  {author} {\bibinfo {author} {\bibfnamefont {R.}~\bibnamefont
  {Portugal}}\ and\ \bibinfo {author} {\bibfnamefont {J.~K.}\ \bibnamefont
  {Moqadam}},\ }\href@noop {} {\bibfield  {journal} {\bibinfo  {journal} {arXiv
  preprint arXiv:2212.08889}\ } (\bibinfo {year} {2022})}\BibitemShut {NoStop}%
\bibitem [{\citenamefont {Kirkpatrick}\ \emph {et~al.}(1983)\citenamefont
  {Kirkpatrick}, \citenamefont {Gelatt~Jr},\ and\ \citenamefont
  {Vecchi}}]{Kirkpatrick1983}%
  \BibitemOpen
  \bibfield  {author} {\bibinfo {author} {\bibfnamefont {S.}~\bibnamefont
  {Kirkpatrick}}, \bibinfo {author} {\bibfnamefont {C.~D.}\ \bibnamefont
  {Gelatt~Jr}}, \ and\ \bibinfo {author} {\bibfnamefont {M.~P.}\ \bibnamefont
  {Vecchi}},\ }\href@noop {} {\bibfield  {journal} {\bibinfo  {journal}
  {Science}\ }\textbf {\bibinfo {volume} {220}},\ \bibinfo {pages} {671}
  (\bibinfo {year} {1983})}\BibitemShut {NoStop}%
\bibitem [{\citenamefont {Ingber}(1993)}]{Ingber1993}%
  \BibitemOpen
  \bibfield  {author} {\bibinfo {author} {\bibfnamefont {L.}~\bibnamefont
  {Ingber}},\ }\href@noop {} {\bibfield  {journal} {\bibinfo  {journal}
  {Mathematical and Computer Modelling}\ }\textbf {\bibinfo {volume} {18}},\
  \bibinfo {pages} {29} (\bibinfo {year} {1993})}\BibitemShut {NoStop}%
\bibitem [{\citenamefont {Short}\ and\ \citenamefont
  {Farrelly}(2012)}]{Short_2012}%
  \BibitemOpen
  \bibfield  {author} {\bibinfo {author} {\bibfnamefont {A.~J.}\ \bibnamefont
  {Short}}\ and\ \bibinfo {author} {\bibfnamefont {T.~C.}\ \bibnamefont
  {Farrelly}},\ }\href@noop {} {\bibfield  {journal} {\bibinfo  {journal} {New
  Journal of Physics}\ }\textbf {\bibinfo {volume} {14}},\ \bibinfo {pages}
  {013063} (\bibinfo {year} {2012})}\BibitemShut {NoStop}%
\bibitem [{\citenamefont {D'Alessio}\ \emph {et~al.}(2016)\citenamefont
  {D'Alessio}, \citenamefont {Kafri}, \citenamefont {Polkovnikov},\ and\
  \citenamefont {Rigol}}]{DAlessio_2015qtq}%
  \BibitemOpen
  \bibfield  {author} {\bibinfo {author} {\bibfnamefont {L.}~\bibnamefont
  {D'Alessio}}, \bibinfo {author} {\bibfnamefont {Y.}~\bibnamefont {Kafri}},
  \bibinfo {author} {\bibfnamefont {A.}~\bibnamefont {Polkovnikov}}, \ and\
  \bibinfo {author} {\bibfnamefont {M.}~\bibnamefont {Rigol}},\ }\href@noop {}
  {\bibfield  {journal} {\bibinfo  {journal} {Advances in Physics}\ }\textbf
  {\bibinfo {volume} {65}},\ \bibinfo {pages} {239} (\bibinfo {year}
  {2016})}\BibitemShut {NoStop}%
\bibitem [{\citenamefont {Deutsch}(2018)}]{Deutsch_2018}%
  \BibitemOpen
  \bibfield  {author} {\bibinfo {author} {\bibfnamefont {J.~M.}\ \bibnamefont
  {Deutsch}},\ }\href@noop {} {\bibfield  {journal} {\bibinfo  {journal}
  {Reports on Progress in Physics}\ }\textbf {\bibinfo {volume} {81}},\
  \bibinfo {pages} {082001} (\bibinfo {year} {2018})}\BibitemShut {NoStop}%
\bibitem [{\citenamefont {Farhi}\ and\ \citenamefont
  {Gutmann}(1998{\natexlab{b}})}]{Farhi_1998}%
  \BibitemOpen
  \bibfield  {author} {\bibinfo {author} {\bibfnamefont {E.}~\bibnamefont
  {Farhi}}\ and\ \bibinfo {author} {\bibfnamefont {S.}~\bibnamefont
  {Gutmann}},\ }\href@noop {} {\bibfield  {journal} {\bibinfo  {journal}
  {Physical Review A}\ }\textbf {\bibinfo {volume} {58}},\ \bibinfo {pages}
  {915–928} (\bibinfo {year} {1998}{\natexlab{b}})}\BibitemShut {NoStop}%
\bibitem [{\citenamefont {Childs}(2010)}]{Childs_2010}%
  \BibitemOpen
  \bibfield  {author} {\bibinfo {author} {\bibfnamefont {A.~M.}\ \bibnamefont
  {Childs}},\ }\href@noop {} {\bibfield  {journal} {\bibinfo  {journal}
  {Communications in Mathematical Physics}\ }\textbf {\bibinfo {volume}
  {294}},\ \bibinfo {pages} {581} (\bibinfo {year} {2010})}\BibitemShut
  {NoStop}%
\bibitem [{\citenamefont {Gogolin}\ and\ \citenamefont
  {Eisert}(2016)}]{Gogolin_2016}%
  \BibitemOpen
  \bibfield  {author} {\bibinfo {author} {\bibfnamefont {C.}~\bibnamefont
  {Gogolin}}\ and\ \bibinfo {author} {\bibfnamefont {J.}~\bibnamefont
  {Eisert}},\ }\href@noop {} {\bibfield  {journal} {\bibinfo  {journal}
  {Reports on Progress in Physics}\ }\textbf {\bibinfo {volume} {79}},\
  \bibinfo {pages} {056001} (\bibinfo {year} {2016})}\BibitemShut {NoStop}%
\bibitem [{\citenamefont {Trushechkin}\ \emph {et~al.}(2022)\citenamefont
  {Trushechkin}, \citenamefont {Merkli}, \citenamefont {Cresser},\ and\
  \citenamefont {Anders}}]{Trushechkin_2022}%
  \BibitemOpen
  \bibfield  {author} {\bibinfo {author} {\bibfnamefont {A.~S.}\ \bibnamefont
  {Trushechkin}}, \bibinfo {author} {\bibfnamefont {M.}~\bibnamefont {Merkli}},
  \bibinfo {author} {\bibfnamefont {J.~D.}\ \bibnamefont {Cresser}}, \ and\
  \bibinfo {author} {\bibfnamefont {J.}~\bibnamefont {Anders}},\ }\href
  {\doibase 10.1116/5.0073853} {\bibfield  {journal} {\bibinfo  {journal}
  {{AVS} Quantum Science}\ }\textbf {\bibinfo {volume} {4}},\ \bibinfo {pages}
  {012301} (\bibinfo {year} {2022})}\BibitemShut {NoStop}%
\bibitem [{\citenamefont {{Ueda}}(2020)}]{QuantumET}%
  \BibitemOpen
  \bibfield  {author} {\bibinfo {author} {\bibfnamefont {M.}~\bibnamefont
  {{Ueda}}},\ }\href {\doibase 10.1038/s42254-020-0237-x} {\bibfield  {journal}
  {\bibinfo  {journal} {Nature Reviews Physics}\ }\textbf {\bibinfo {volume}
  {2}},\ \bibinfo {pages} {669} (\bibinfo {year} {2020})}\BibitemShut {NoStop}%
\bibitem [{\citenamefont {Schr{\"o}dinger}(1989)}]{schrodinger1989statistical}%
  \BibitemOpen
  \bibfield  {author} {\bibinfo {author} {\bibfnamefont {E.}~\bibnamefont
  {Schr{\"o}dinger}},\ }\href@noop {} {\emph {\bibinfo {title} {Statistical
  thermodynamics}}}\ (\bibinfo  {publisher} {Courier Corporation},\ \bibinfo
  {year} {1989})\BibitemShut {NoStop}%
\bibitem [{\citenamefont {Gemmer}\ and\ \citenamefont
  {Mahler}(2003)}]{gemmer2003distribution}%
  \BibitemOpen
  \bibfield  {author} {\bibinfo {author} {\bibfnamefont {J.}~\bibnamefont
  {Gemmer}}\ and\ \bibinfo {author} {\bibfnamefont {G.}~\bibnamefont
  {Mahler}},\ }\href@noop {} {\bibfield  {journal} {\bibinfo  {journal} {The
  European Physical Journal B-Condensed Matter and Complex Systems}\ }\textbf
  {\bibinfo {volume} {31}},\ \bibinfo {pages} {249} (\bibinfo {year}
  {2003})}\BibitemShut {NoStop}%
\bibitem [{\citenamefont {Goldstein}\ \emph {et~al.}(2006)\citenamefont
  {Goldstein}, \citenamefont {Lebowitz}, \citenamefont {Tumulka},\ and\
  \citenamefont {Zangh{\`\i}}}]{Goldstein_2006}%
  \BibitemOpen
  \bibfield  {author} {\bibinfo {author} {\bibfnamefont {S.}~\bibnamefont
  {Goldstein}}, \bibinfo {author} {\bibfnamefont {J.~L.}\ \bibnamefont
  {Lebowitz}}, \bibinfo {author} {\bibfnamefont {R.}~\bibnamefont {Tumulka}}, \
  and\ \bibinfo {author} {\bibfnamefont {N.}~\bibnamefont {Zangh{\`\i}}},\
  }\href@noop {} {\bibfield  {journal} {\bibinfo  {journal} {Physical review
  letters}\ }\textbf {\bibinfo {volume} {96}},\ \bibinfo {pages} {050403}
  (\bibinfo {year} {2006})}\BibitemShut {NoStop}%
\bibitem [{\citenamefont {Popescu}\ \emph {et~al.}(2006)\citenamefont
  {Popescu}, \citenamefont {Short},\ and\ \citenamefont
  {Winter}}]{popescu2006entanglement}%
  \BibitemOpen
  \bibfield  {author} {\bibinfo {author} {\bibfnamefont {S.}~\bibnamefont
  {Popescu}}, \bibinfo {author} {\bibfnamefont {A.~J.}\ \bibnamefont {Short}},
  \ and\ \bibinfo {author} {\bibfnamefont {A.}~\bibnamefont {Winter}},\
  }\href@noop {} {\bibfield  {journal} {\bibinfo  {journal} {Nature Physics}\
  }\textbf {\bibinfo {volume} {2}},\ \bibinfo {pages} {754} (\bibinfo {year}
  {2006})}\BibitemShut {NoStop}%
\bibitem [{\citenamefont {Linden}\ \emph {et~al.}(2009)\citenamefont {Linden},
  \citenamefont {Popescu}, \citenamefont {Short},\ and\ \citenamefont
  {Winter}}]{Linden_2009}%
  \BibitemOpen
  \bibfield  {author} {\bibinfo {author} {\bibfnamefont {N.}~\bibnamefont
  {Linden}}, \bibinfo {author} {\bibfnamefont {S.}~\bibnamefont {Popescu}},
  \bibinfo {author} {\bibfnamefont {A.~J.}\ \bibnamefont {Short}}, \ and\
  \bibinfo {author} {\bibfnamefont {A.}~\bibnamefont {Winter}},\ }\href@noop {}
  {\bibfield  {journal} {\bibinfo  {journal} {Physical Review E}\ }\textbf
  {\bibinfo {volume} {79}} (\bibinfo {year} {2009})}\BibitemShut {NoStop}%
\bibitem [{\citenamefont {Dahlsten}\ \emph {et~al.}(2014)\citenamefont
  {Dahlsten}, \citenamefont {Lupo}, \citenamefont {Mancini},\ and\
  \citenamefont {Serafini}}]{Dahlsten_2014}%
  \BibitemOpen
  \bibfield  {author} {\bibinfo {author} {\bibfnamefont {O.~C.~O.}\
  \bibnamefont {Dahlsten}}, \bibinfo {author} {\bibfnamefont {C.}~\bibnamefont
  {Lupo}}, \bibinfo {author} {\bibfnamefont {S.}~\bibnamefont {Mancini}}, \
  and\ \bibinfo {author} {\bibfnamefont {A.}~\bibnamefont {Serafini}},\
  }\href@noop {} {\bibfield  {journal} {\bibinfo  {journal} {Journal of Physics
  A: Mathematical and Theoretical}\ }\textbf {\bibinfo {volume} {47}},\
  \bibinfo {pages} {363001} (\bibinfo {year} {2014})}\BibitemShut {NoStop}%
\bibitem [{\citenamefont {M{\"u}ller}\ \emph {et~al.}(2012)\citenamefont
  {M{\"u}ller}, \citenamefont {Dahlsten},\ and\ \citenamefont
  {Vedral}}]{muller2012unifying}%
  \BibitemOpen
  \bibfield  {author} {\bibinfo {author} {\bibfnamefont {M.~P.}\ \bibnamefont
  {M{\"u}ller}}, \bibinfo {author} {\bibfnamefont {O.~C.}\ \bibnamefont
  {Dahlsten}}, \ and\ \bibinfo {author} {\bibfnamefont {V.}~\bibnamefont
  {Vedral}},\ }\href@noop {} {\bibfield  {journal} {\bibinfo  {journal}
  {Communications in Mathematical Physics}\ }\textbf {\bibinfo {volume}
  {316}},\ \bibinfo {pages} {441} (\bibinfo {year} {2012})}\BibitemShut
  {NoStop}%
\bibitem [{\citenamefont {Lostaglio}\ \emph {et~al.}(2015)\citenamefont
  {Lostaglio}, \citenamefont {Korzekwa}, \citenamefont {Jennings},\ and\
  \citenamefont {Rudolph}}]{Matteo}%
  \BibitemOpen
  \bibfield  {author} {\bibinfo {author} {\bibfnamefont {M.}~\bibnamefont
  {Lostaglio}}, \bibinfo {author} {\bibfnamefont {K.}~\bibnamefont {Korzekwa}},
  \bibinfo {author} {\bibfnamefont {D.}~\bibnamefont {Jennings}}, \ and\
  \bibinfo {author} {\bibfnamefont {T.}~\bibnamefont {Rudolph}},\ }\href@noop
  {} {\bibfield  {journal} {\bibinfo  {journal} {Physical Review X}\ }\textbf
  {\bibinfo {volume} {5}} (\bibinfo {year} {2015})}\BibitemShut {NoStop}%
\bibitem [{\citenamefont {Reimann}(2008)}]{Reimann}%
  \BibitemOpen
  \bibfield  {author} {\bibinfo {author} {\bibfnamefont {P.}~\bibnamefont
  {Reimann}},\ }\href {\doibase 10.1103/PhysRevLett.101.190403} {\bibfield
  {journal} {\bibinfo  {journal} {Phys. Rev. Lett.}\ }\textbf {\bibinfo
  {volume} {101}},\ \bibinfo {pages} {190403} (\bibinfo {year}
  {2008})}\BibitemShut {NoStop}%
\bibitem [{\citenamefont {Reimann}(2010)}]{reimann2010canonical}%
  \BibitemOpen
  \bibfield  {author} {\bibinfo {author} {\bibfnamefont {P.}~\bibnamefont
  {Reimann}},\ }\href@noop {} {\bibfield  {journal} {\bibinfo  {journal} {New
  Journal of Physics}\ }\textbf {\bibinfo {volume} {12}},\ \bibinfo {pages}
  {055027} (\bibinfo {year} {2010})}\BibitemShut {NoStop}%
\bibitem [{\citenamefont {Chung}\ and\ \citenamefont
  {Sternberg}(1993)}]{bucky1}%
  \BibitemOpen
  \bibfield  {author} {\bibinfo {author} {\bibfnamefont {F.}~\bibnamefont
  {Chung}}\ and\ \bibinfo {author} {\bibfnamefont {S.}~\bibnamefont
  {Sternberg}},\ }\href {http://www.jstor.org/stable/29774821} {\bibfield
  {journal} {\bibinfo  {journal} {American Scientist}\ }\textbf {\bibinfo
  {volume} {81}},\ \bibinfo {pages} {56} (\bibinfo {year} {1993})}\BibitemShut
  {NoStop}%
\bibitem [{\citenamefont {Grünbaum}\ and\ \citenamefont {Motzkin}(1963)}]{gm}%
  \BibitemOpen
  \bibfield  {author} {\bibinfo {author} {\bibfnamefont {B.}~\bibnamefont
  {Grünbaum}}\ and\ \bibinfo {author} {\bibfnamefont {T.~S.}\ \bibnamefont
  {Motzkin}},\ }\href@noop {} {\bibfield  {journal} {\bibinfo  {journal}
  {Canadian Journal of Mathematics}\ }\textbf {\bibinfo {volume} {15}},\
  \bibinfo {pages} {744–751} (\bibinfo {year} {1963})}\BibitemShut {NoStop}%
\bibitem [{\citenamefont {Brinkmann}\ \emph {et~al.}(2013)\citenamefont
  {Brinkmann}, \citenamefont {Coolsaet}, \citenamefont {Goedgebeur},\ and\
  \citenamefont {Mélot}}]{BRINKMANN}%
  \BibitemOpen
  \bibfield  {author} {\bibinfo {author} {\bibfnamefont {G.}~\bibnamefont
  {Brinkmann}}, \bibinfo {author} {\bibfnamefont {K.}~\bibnamefont {Coolsaet}},
  \bibinfo {author} {\bibfnamefont {J.}~\bibnamefont {Goedgebeur}}, \ and\
  \bibinfo {author} {\bibfnamefont {H.}~\bibnamefont {Mélot}},\ }\href@noop {}
  {\bibfield  {journal} {\bibinfo  {journal} {Discrete Applied Mathematics}\
  }\textbf {\bibinfo {volume} {161}},\ \bibinfo {pages} {311} (\bibinfo {year}
  {2013})}\BibitemShut {NoStop}%
\bibitem [{\citenamefont {Dunlop}\ \emph {et~al.}(2021)\citenamefont {Dunlop},
  \citenamefont {Cohen},\ and\ \citenamefont {Short}}]{Dunlop_2021}%
  \BibitemOpen
  \bibfield  {author} {\bibinfo {author} {\bibfnamefont {J.}~\bibnamefont
  {Dunlop}}, \bibinfo {author} {\bibfnamefont {O.}~\bibnamefont {Cohen}}, \
  and\ \bibinfo {author} {\bibfnamefont {A.~J.}\ \bibnamefont {Short}},\
  }\href@noop {} {\bibfield  {journal} {\bibinfo  {journal} {Physical Review
  E}\ }\textbf {\bibinfo {volume} {104}},\ \bibinfo {pages} {024135} (\bibinfo
  {year} {2021})}\BibitemShut {NoStop}%
\bibitem [{\citenamefont {Grover}(2001)}]{grover2001schrodinger}%
  \BibitemOpen
  \bibfield  {author} {\bibinfo {author} {\bibfnamefont {L.~K.}\ \bibnamefont
  {Grover}},\ }\href@noop {} {\bibfield  {journal} {\bibinfo  {journal}
  {Pramana}\ }\textbf {\bibinfo {volume} {56}},\ \bibinfo {pages} {333}
  (\bibinfo {year} {2001})}\BibitemShut {NoStop}%
\bibitem [{\citenamefont {Diaconis}(2005)}]{Diaconis}%
  \BibitemOpen
  \bibfield  {author} {\bibinfo {author} {\bibfnamefont {P.}~\bibnamefont
  {Diaconis}},\ }\href@noop {} {\bibfield  {journal} {\bibinfo  {journal}
  {AMS}\ }\textbf {\bibinfo {volume} {52}},\ \bibinfo {pages} {1348} (\bibinfo
  {year} {2005})}\BibitemShut {NoStop}%
\bibitem [{\citenamefont {Peres}(1995)}]{Peres}%
  \BibitemOpen
  \bibfield  {author} {\bibinfo {author} {\bibfnamefont {A.}~\bibnamefont
  {Peres}},\ }\href {https://books.google.co.in/books?id=rMGqMyFBcL8C} {\emph
  {\bibinfo {title} {Quantum Theory: Concepts and Methods}}},\ Fundamental
  Theories of Physics\ (\bibinfo  {publisher} {Springer Netherlands},\ \bibinfo
  {year} {1995})\BibitemShut {NoStop}%
\bibitem [{\citenamefont {Lee}\ \emph {et~al.}(1992)\citenamefont {Lee},
  \citenamefont {Luo}, \citenamefont {Sagan},\ and\ \citenamefont
  {Yeh}}]{lee1992eigenvector}%
  \BibitemOpen
  \bibfield  {author} {\bibinfo {author} {\bibfnamefont {S.-L.}\ \bibnamefont
  {Lee}}, \bibinfo {author} {\bibfnamefont {Y.-L.}\ \bibnamefont {Luo}},
  \bibinfo {author} {\bibfnamefont {B.~E.}\ \bibnamefont {Sagan}}, \ and\
  \bibinfo {author} {\bibfnamefont {Y.-N.}\ \bibnamefont {Yeh}},\ }\href@noop
  {} {\bibfield  {journal} {\bibinfo  {journal} {International journal of
  quantum chemistry}\ }\textbf {\bibinfo {volume} {41}},\ \bibinfo {pages}
  {105} (\bibinfo {year} {1992})}\BibitemShut {NoStop}%
\bibitem [{\citenamefont {Cantoni}\ and\ \citenamefont
  {Butler}(1976)}]{cantoni1976eigenvalues}%
  \BibitemOpen
  \bibfield  {author} {\bibinfo {author} {\bibfnamefont {A.}~\bibnamefont
  {Cantoni}}\ and\ \bibinfo {author} {\bibfnamefont {P.}~\bibnamefont
  {Butler}},\ }\href {\doibase 10.1016/0024-3795(76)90101-4} {\bibfield
  {journal} {\bibinfo  {journal} {Linear Algebra and its Applications}\
  }\textbf {\bibinfo {volume} {13}},\ \bibinfo {pages} {275} (\bibinfo {year}
  {1976})}\BibitemShut {NoStop}%
\end{thebibliography}%

\appendix

\section{Definitions and derivations}\label{ddp}
\noindent {\bf {\em $\Tilde{H}_B$ and $\Tilde{H}_{int}$ definition---}}
We define $\Tilde{H}_B := I_S \otimes H_B$, where

\begin{equation}\label{bath}
    \bra{b}^B_k H_B\ket{b}^B_j = \begin{cases}
       1 &  \ket{b}^B_j \neq \ket{b}^B_k, \mathrm{edge}(j,k) \in F_N\\
        &  j,k \in \{6,7,\dots,N-1,N\}\\
        0     & \text{otherwise},
    \end{cases}
\end{equation}

for $\mathcal{H}_B = \mathrm{span}\{\ket{b}^B_j \vert 6 \leq j \leq N \}$. 

\begin{equation}\label{eq:nowalkerbath}
 \begin{aligned}
   \bra{b}^B_0 H_B\ket{b}^B_0  & = 0\\
   \bra{b}^B_0 H_B\ket{b}^B_j  & = 0\\
   \bra{b}^B_k H_B\ket{b}^B_0  & = 0
\end{aligned}
\end{equation}
for $j,k \in \{6,7,\dots,N-1,N\}$.

Finally, the remaining part of the Hamiltonian is the interaction Hamiltonian $\Tilde H_{int}$, defined via

\begin{equation}\label{interaction}
    \bra{b}_k \Tilde H_{int}\ket{b}_j = \begin{cases}
       1 &  \ket{b}_j \neq \ket{b}_k, \mathrm{edge}(j,k) \in F_N\\
        &  j,k \in \{1,2,\dots,9,10\} \\
       0     & \text{otherwise}.
    \end{cases}
\end{equation}

\begin{equation}\label{eq:nowalkerint}
 \begin{aligned}
   \bra{b}_0 \Tilde H_{int}\ket{b}_0  & = 0\\
   \bra{b}_0 \Tilde H_{int}\ket{b}_j  & = 0\\
   \bra{b}_k \Tilde H_{int}\ket{b}_0  & = 0
\end{aligned}   
\end{equation}
for $j,k \in \{1,2,\dots,9,10\}$.

\noindent {\bf {\em Pentagon Gibbs state derivation---}}
We used the fact that $\braket{j}{\lambda_j}\braket{\lambda_j}{j} = 1/5$. As $\beta \rightarrow \infty$,
\begin{align*}
    \lim_{\beta \rightarrow \infty} p(j) & = \lim_{\beta \rightarrow \infty} \frac{1}{Z}\sum_{j = 0}^{4}\frac{\exp{(-\beta \cos{(2 \pi j/5)})}}{5}\\
    & = \frac{1}{5}\lim_{\beta \rightarrow \infty} \frac{e^{-\beta} + 2 e^{((1 + \sqrt{5})\beta/4)} + 2 e^{(-(\sqrt{5}-1)\beta/4)}}{e^{-\beta} + 2 e^{((1 + \sqrt{5})\beta/4)} + 2 e^{(-(\sqrt{5}-1)\beta/4)} + 1}\\
    & = \frac{1}{5}.
\end{align*}

In the other extreme, as $\beta \rightarrow 0 $, $\lim_{\beta \rightarrow 0 } \frac{1}{Z} = \frac{1}{6}$, and 
\begin{align*}
    \lim_{\beta \rightarrow 0} p(j) & = \lim_{\beta \rightarrow 0} \frac{1}{Z}\sum_{j = 0}^{4}\frac{\exp{(-\beta \cos{(2 \pi j/5)})}}{5}\\
    & = \frac{1}{Z}\sum_{l = 0}^{4}\frac{1}{5}\\
    & = \frac{1}{6}\, \forall j\in  \{1,2,3,4,5\}.
\end{align*}

\section{ETH argument for projectors}\label{Eth_o}

\begin{figure}[htbp]
        \centering
         \includegraphics[trim = 90 260 70 235,width=0.55\textwidth,keepaspectratio, left]{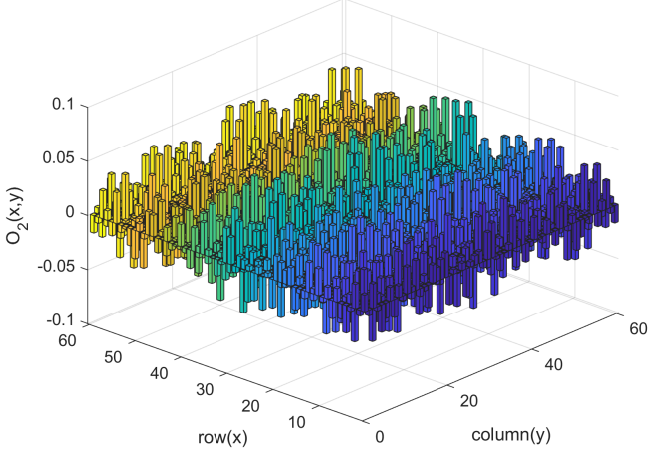}
         \caption{ {{\bf $O_2$ projector in Hamiltonian eigenbasis:}} We plot the matrix elements of $O_2 =\ketbra{2}{2}$ in the energy eigenbasis of $H$, for the C60 graph. Observe that the diagonal of $O$ in energy eigenbasis is a not smooth function at all and off-diagonal terms are also significant. This shows that ETH relation is not respected for this projector. }
        \label{fig:ETHplotO_2}    
    \end{figure}

The ETH relation does not hold for the $O_x = \ketbra{x}{x}$ observables. An example can be seen in Fig.~\ref{fig:ETHplotO_2}.  We numerically find that for $1 \leq x \leq 5$ the diagonal of $O_x$ is not a smooth function, varying wildly between different increments of $m$. Denoting $\langle O_x^{mm}\rangle =\frac{\sum_m O_x^{mm}}{60}$, we find that for the 5 nodes in the pentagon of interest, $\langle O_1^{mm}\rangle=0.017\pm
0.028$, $\langle O_2^{mm}\rangle=0.017\pm
0.018$, $\langle O_3^{mm}\rangle=0.017\pm
0.018$, $\langle O_4^{mm}\rangle=0.017\pm
0.015$, $\langle O_5^{mm}\rangle=0.017\pm
0.017$, where the number after $\pm$ is the standard deviation. In this case it seems natural to conclude that ETH does not hold, since the diagonal fluctuates such that different neighbouring $m's$ can have different $O_2^{m m}$. In particular,  $\langle O_2 \rangle$ will depend heavily on which nearby $m's$ the initial state has support on, such that Eq.~\ref{eq:equalsmc} would be violated even for narrow initial energy windows. 

\begin{figure}[htbp]
        \centering
        \includegraphics[trim = 90 260 70 235,width=0.55\textwidth,keepaspectratio, left]{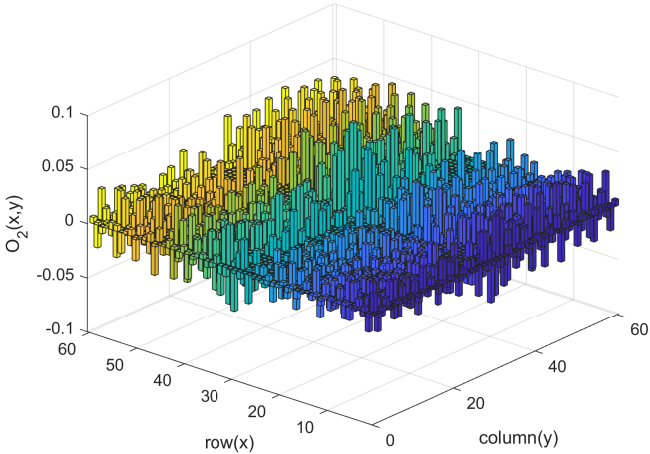}
        \caption{ {{\bf $O_2 = \ketbra{2}{2}$ projector in random basis(real):}} We plot the matrix elements of $O_2 =\ketbra{2}{2}$ in the random real basis. Observe that the diagonal of $O$ in the real random basis is a not smooth function at all and off-diagonal terms are also significant. This shows that ETH relation is not respected for this projector. }
        \label{fig:ETHplotO_r}    
    \end{figure}

To partially explain the form of the node projectors in the energy eigenbasis, we note that there are similarities between how the node states look in the energy eigenbasis with how a random state would look. Here a random state means a Haar random state, with respect to the orthogonal or unitary groups~\cite{Diaconis}. It is natural to consider the orthogonal group since the position nodes have real entries in the energy eigenbasis for quantum walks on graphs wherein the Hamiltonian is real in the node basis. Fig.~\ref{fig:ETHplotO_2} and Fig.~\ref{fig:ETHplotO_r} show the qualitative similarity. To understand how a projector looks in a random basis, consider $O_2 = \ketbra{2}{2}$. Then $O_2^{mn} =\braket{m}{2}\braket{2}{n} = \psi_2^{m*} \psi_2^{n}$. If the $\psi_2^{m}$ are real entries picked from the Haar measure then the (non-negative) diagonal entries $|\psi_2^{m}|^2$ are fluctuating and sum to 1, and the off-diagonal entries are also fluctuating with similar absolute value size to the diagonal (as depicted in Fig.~\ref{fig:ETHplotO_r}). 

Another way to argue for a similarity between the case of the fullerene energy eigenbasis and a random basis is to compare the respective measurement entropies of a state for all node states in relation to the eigenstates of buckyball Hamiltonian~\cite{Peres}: $E = - \sum_{k} p_k \log{p_k}$, where $p_k = |\braket{\lambda_k}{x}|^2$ for all $1 \leq x \leq 60$ and $H = \sum_{k} \lambda_k \ketbra{\lambda_k}{\lambda_k}$. In this way, we can quantify how many eigenstates of $H$ are overlapping with a given node state. We find the expectation value $\langle E \rangle=3.57\pm 0.16$ for the buckyball node states. For comparison, we picked sixty random states with respect to orthogonal group of the same dimension and their average energy measurement entropy $\langle E \rangle=3.39 \pm 0.10$, which is not identical but close.  


\section{Microcanonical state vs. Boltzmann distribution}\label{microcanonical}
For completeness, we quickly summarise a standard argument concerning the relation between (i) the Boltzmann/Gibbs distribution and (ii) the so-called microcanonical ensemble. (We write this in terms of classical probability distributions which could equivalently have been written as diagonal quantum density matrices in the energy eigenbasis). More specifically, (i) means $p(a)=\exp(-\beta E_a)/Z$ where a is a state on the system A, with energy $E_a$ (which is thus assumed to be well-defined) and $Z$ is a normalisation factor. (ii) involves including an environment with state $b$ and energy $E_b$ and demanding that $p(a,b)=u$ for $E_a+E_b=E_{tot}$, and 0 else, where $u$ is a constant. (ii) implies (i) (up to $c=\beta$) if we, crucially, assume that $p(b|a)=\mathcal {N}\exp(-cE_b)$ for $E_a\leq E_{tot}$, where $\mathcal {N}$ is a normalisation constant. This assumption is consistent with assuming that the number of states on $b$ for a given energy grows exponentially and that they are all equally likely for a given $E_b$. To finalise the argument, recall
$p(b|a):=p(a,b)/p(a)$ s.t.\ $p(a)=p(a,b)/p(b|a)=\frac{u}{\mathcal{N}}\exp(c(E_b))=\frac{u}{\mathcal{N}}\exp(c(E_{tot}-E_a))=\exp(-cE_a)(\frac{u}{\mathcal{N}}\exp(cE_{tot})):=\exp(-cE_a)\mathcal{M}$, for $E_A\leq E_{tot}$, and 0 else.

To relate $c$ to $\beta:=\frac{1}{k_BT}$, one may demand consistency with other definitions of temperature $T$ and its relation to entropy. For example consider demanding that adding a little energy $d\langle  E_a \rangle$ to a thermalised system, should change its thermodynamic entropy $S=-k_B\sum_i p_i\ln p_i$ by 
\begin{equation}
\label{eq:dSdemand}
dS=\frac{d\langle  E_a \rangle}{T},
\end{equation}
where $\langle  E_a \rangle:=\sum_a p_a E_a $. We demand, as is standard in stochastic thermodynamics (recall $dU=\sum_a d(p_a)E_a+\sum_a p_adE_a\equiv dQ-dW$) that heating  means the system's energy eigenstates are invariant and only their probabilities change, s.t.\  $d\langle  E_a \rangle=\sum_a dp_a E_a$. Then
\begin{eqnarray*}
dS &=& k_B\sum_a d\left(p_a\ln \frac{1}{p_a}\right)\\
   &=& k_B\sum_a \left(d(p_a)\ln \frac{1}{p_a}+p_a d\left(\ln  \frac{1}{p_a}\right)\right)\\
   &=& k_B\sum_a \left(d(p_a)\ln \frac{1}{p_a}+p_a  \frac{dp_a}{p_a}\right)\\
   &=& - k_B\sum_a d(p_a)\ln (\exp(-cE_a)\mathcal{M})\\
   &=& - k_B\sum_a d(p_a)(-cE_a +\ln \mathcal{M}) \\
   &=&k_B\sum_a d(p_a) c E_a - k_B \sum_a d(p_a) \ln \mathcal{M}\\
   &=&k_B c \sum_a d(p_a) E_a\\
   &=&k_B c (d\langle  E_a \rangle).
\end{eqnarray*}
We used $\sum_a dp_a=0$ which follows from $\sum_a p_a=1$. The above expression for $dS$ implies, when demanding Eq.~\ref{eq:dSdemand}, that  $k_Bc=\frac{1}{T}$ and $c=\frac{1}{k_BT}=\beta$.

\section{Proof of equation \ref{equal}}

The quantum walk limiting distribution on $C60$ is depicted in FIG.~\ref{fig:buckyball3d}. It exhibits a distinctive cross-like pattern which emerges due to the radial symmetry present in the eigenvectors $\ket{\lambda_k}$ of the adjacency matrix $A$, i.e. 

\begin{equation}\label{equal2}
    \braket{x}{\lambda_k} = \pm \braket{61-x}{\lambda_k},
\end{equation}
where $1 \leq x \leq 60$. 

The overview of the argument is as follows. We define a matrix $A_2$ for which it is easier to find the eigenvectors than it is for $A$ (because of $A_2$ having a block structure), and which has the same eigenvectors as $A$. We then find the eigenvectors of $A_2$. Finally, we show that these eigenvectors have the symmetry of Eq.~\ref{equal2}. 

We shall make use of results from Ref.~\cite{lee1992eigenvector}. The adjacency matrix given in Ref.~\cite{lee1992eigenvector}, which we now call $A_1$ is the same as our $A$ here up to permutations. By inspection $P^{-1}A P = A_1$ for some permutation matrix $P$. Since permutation matrices are unitary, the spectrum of all distinct adjacency matrices generated by the different labelling of vertices of the same graph is the same. We create another adjacency matrix called $A_2$ using $P_1$ (defined in Eq.~\ref{P1} ) from $A_1$ to prove Eq.~\ref{equal2}. Since $A$ and $A_2$ are centrosymmetric matrices the eigenvector distributions of both matrices display a similar pattern. A matrix $S_{n\times n} = [S_{j,k}]$ is centrosymmetric when $S_{j,k} = S_{n-j, n-k+1}$ for $j,k \in \{1,2,\dots, n\}$.     

The adjacency matrix $A_2$ of fullerene $C60$
is $A_2 = P_1^{-1} A_1 P_1$ for $A_1$ given in~\cite{lee1992eigenvector}. $A_2$ is given as follows:

\setcounter{MaxMatrixCols}{12}
\begin{equation}
\begin{bmatrix}
A(C_5) & I & 0 & 0 & 0 & 0 & 0 & 0 & 0 & 0 & 0 & 0 \\
 I & 0 & I & I & 0 & 0 & 0 & 0 & 0 & 0 & 0 & 0 \\
 0 & I & 0 & K^t & 0 & I & 0 & 0 & 0 & 0 & 0 & 0 \\
 0 & I & K & 0 & I & 0 & 0 & 0 & 0 & 0 & 0 & 0 \\ 
 0 & 0 & 0 & I & 0 & I & L^t J & 0 & 0 & 0 & 0 & 0 \\ 
 0 & 0 & I & 0 & I & 0 & 0 & L J & 0 & 0 & 0 & 0 \\ 
 0 & 0 & 0 & 0 & J L & 0 & 0 & I & 0 & I & 0 & 0 \\ 
 0 & 0 & 0 & 0 & 0 & J L^{t} & I & 0 & I & 0 & 0 & 0 \\
  0 & 0 & 0 & 0 & 0 & 0 & 0 & I & 0 & K^t & I & 0 \\
 0 & 0 & 0 & 0 & 0 & 0 & I & 0 & K & 0 & I & 0 \\
 0 & 0 & 0 & 0 & 0 & 0 & 0 & 0 & I & I & 0 & I \\
  0 & 0 & 0 & 0 & 0 & 0 & 0 & 0 & 0 & 0 & I & A(C_5) \\
 \end{bmatrix}, 
\end{equation}

\setcounter{MaxMatrixCols}{12}
\begin{equation}\label{P1}
P_1 = \begin{bmatrix}
I & 0 & 0 & 0 & 0 & 0 & 0 & 0 & 0 & 0 & 0 & 0 \\
 0 & I & 0 & 0 & 0 & 0 & 0 & 0 & 0 & 0 & 0 & 0 \\
 0 & 0 & I & 0 & 0 & 0 & 0 & 0 & 0 & 0 & 0 & 0 \\
 0 & 0 & 0 & I & 0 & 0 & 0 & 0 & 0 & 0 & 0 & 0 \\ 
 0 & 0 & 0 & 0 & I & 0 & 0 & 0 & 0 & 0 & 0 & 0 \\ 
 0 & 0 & 0 & 0 & 0 & I & 0 & 0 & 0 & 0 & 0 & 0 \\ 
 0 & 0 & 0 & 0 & 0 & 0 & J & 0 & 0 & 0 & 0 & 0 \\ 
 0 & 0 & 0 & 0 & 0 & 0 & 0 & J& 0 & 0 & 0 & 0 \\
  0 & 0 & 0 & 0 & 0 & 0 & 0 & 0 & J & 0 & 0 & 0 \\
 0 & 0 & 0 & 0 & 0 & 0 & 0 & 0 & 0 & J & 0 & 0 \\
 0 & 0 & 0 & 0 & 0 & 0 & 0 & 0 & 0 & 0 & J & 0 \\
  0 & 0 & 0 & 0 & 0 & 0 & 0 & 0 & 0 & 0 & 0 & J \\
 \end{bmatrix}, 
\end{equation} 

 where $I$ is the $5 \times 5$ identity matrix and 
\begin{equation}
 J = 
\begin{bmatrix}
  0 & 0 & 0 & 0 & 1 \\ 
  0 & 0 & 0 & 1 & 0  \\
 0 & 0 & 1 & 0 & 0 \\
 0 & 1 & 0 & 0 & 0  \\
  1 & 0 & 0 & 0 & 0  
 \end{bmatrix}.    
\end{equation}

The matrices $A(C_5)$, $K$, and $L$ in $A_2$ are circulant matrices. A circulant matrix is defined as follows: Let $C$ be a circulant matrix, then it has the form:
\begin{equation}
C = \begin{bmatrix}
c_1 & c_2 & \dots & c_n\\
c_n & c_1 & \dots & c_{n-1} \\
c_{n-1} & c_n & \dots & c_{n-2}\\
\dots & \dots & \dots & \dots\\
c_2 & c_3 & \dots & c_1
\end{bmatrix}.
\end{equation}

$C$ can be denoted as $C = [[c_1, c_2,  \dots , c_n]]$. Similarly, we write 
\[A(C_5)=[[0, 1, 0, 0, 1 ]], \\
K =[[0, 1, 0, 0, 0 ]],\\
L  =[[0, 0, 1, 0, 0 ]] .\]

Now we write $A_2$ again in blocks as

\begin{equation}
    A_2 =\begin{bmatrix}
    B & \mathcal{J}C\mathcal{J}\\
    C & \mathcal{J}B\mathcal{J}
\end{bmatrix},
\end{equation}

where \begin{equation}
    B = 
\begin{bmatrix}
A(C_5) & I & 0 & 0 & 0 & 0 \\
 I & 0 & I & I & 0 & 0 \\
 0 & I & 0 & K^t & 0 & I \\
 0 & I & K & 0 & I & 0 \\ 
 0 & 0 & 0 & I & 0 & I \\ 
 0 & 0 & I & 0 & I & 0 
 \end{bmatrix},
\end{equation}

\begin{equation}
    C = 
\begin{bmatrix}
 0 & 0 & 0 & 0 & JL & 0 \\ 
 0 & 0 & 0 & 0 & 0 & JL^{t}  \\
  0 & 0 & 0 & 0 & 0 & 0 \\
 0 & 0 & 0 & 0 & 0 & 0  \\
 0 & 0 & 0 & 0 & 0 & 0 \\
  0 & 0 & 0 & 0 & 0 & 0 
 \end{bmatrix}, 
\end{equation}
and
\begin{equation}
    \mathcal{J} = 
\begin{bmatrix}
 0 & 0 & 0 & 0 & 0 & J \\ 
 0 & 0 & 0 & 0 & J & 0  \\
  0 & 0 & 0 & J & 0 & 0 \\
 0 & 0 & J & 0 & 0 & 0  \\
 0 & J & 0 & 0 & 0 & 0 \\
  J & 0 & 0 & 0 & 0 & 0 
 \end{bmatrix}.
\end{equation}

Following the reasoning of the proof provided in Ref.~\cite{cantoni1976eigenvalues} concerning centrosymmetric matrices, let us define a matrix

\begin{equation}
    R = \frac{1}{\sqrt{2}}\begin{bmatrix}
    I & -\mathcal{J}\\
    I & \mathcal{J}
\end{bmatrix}.
\end{equation}

Then 
\begin{align*}
    RA &= \frac{1}{\sqrt{2}}\begin{bmatrix}
    I & -\mathcal{J}\\
    I & \mathcal{J}
\end{bmatrix}\begin{bmatrix}
    B & \mathcal{J}C\mathcal{J}\\
    C & \mathcal{J}B\mathcal{J}
\end{bmatrix} \\
&= \frac{1}{\sqrt{2}}\begin{bmatrix}
    B - \mathcal{J}C & \mathcal{J}C\mathcal{J} -  B\mathcal{J} \\
    B + \mathcal{J}C & \mathcal{J}C\mathcal{J} +  B \mathcal{J}
\end{bmatrix}.
\end{align*}

\begin{align*}
    RAR^T &= \frac{1}{2}\begin{bmatrix}
    B - \mathcal{J}C & \mathcal{J}C\mathcal{J} -  B\mathcal{J} \\
    B + \mathcal{J}C & \mathcal{J}C\mathcal{J} +  B\mathcal{J}
\end{bmatrix} \begin{bmatrix}
    I & I\\
    -\mathcal{J} & \mathcal{J}
\end{bmatrix} \\
 &=  \begin{bmatrix}
    B -\mathcal{J}C & 0 \\
    0 & B+\mathcal{J}C
\end{bmatrix}.
\end{align*}

Let us consider the sets of orthogonal eigenvectors $U$ and $V$ for $B - \mathcal{J}C$ and $B + \mathcal{J}C$, respectively. We have

\begin{equation}
    U^T(B -\mathcal{J}C) U = D_1,
\end{equation}
\begin{equation}
 V^T(B +\mathcal{J}C) V = D_2,   
\end{equation}

where $D_1$ and $D_2$ are diagonal matrices with eigenvalues $(B -\mathcal{J}C)$ and $(B +\mathcal{J}C)$ respectively. 

This implies 
\begin{align*}
    M^T RAR^T M &= \begin{bmatrix}
    U^T & 0\\
    0 & V^T
\end{bmatrix}  \begin{bmatrix}
    B -\mathcal{J}C & 0 \\
    0 & B+\mathcal{J}C
\end{bmatrix} \begin{bmatrix}
    U & 0\\
    0 & V
\end{bmatrix}\\
&= \begin{bmatrix}
    D_1 & 0 \\
    0 & D_2
\end{bmatrix},
\end{align*}

for 

\begin{equation}
  M = \begin{bmatrix}
    U & 0\\
    0 & V
\end{bmatrix}.  
\end{equation}

This tells us that $R^T M$ is an eigenvector matrix for $A_2$ where
\begin{align*}
    R^T M &= \frac{1}{\sqrt{2}} \begin{bmatrix}
    I & I\\
    -\mathcal{J} & \mathcal{J}
\end{bmatrix} \begin{bmatrix}
    U & 0\\
    0 & V
\end{bmatrix}\\
&=  \frac{1}{\sqrt{2}}\begin{bmatrix}
    U & V\\
    -\mathcal{J} U  & \mathcal{J} V
\end{bmatrix}.
\end{align*}

Suppose $U = [\ket{u_1}, \ket{u_2}, \ket{u_3}, \dots, \ket{u_{30}}]$, $V = [\ket{v_{1}}, \ket{v_{1}},\ket{v_{3}},\dots,\ket{v_{30}}]$. The eigenvectors of $A_2$ are $W = [\ket{\lambda_1}, \ket{\lambda_2}, \dots, \ket{\lambda_{60}}]$, where

\[\ket{\lambda_j} = \frac{1}{\sqrt{2}}\begin{bmatrix}
    \ket{u_j}\\
    -\mathcal{J} \ket{u_j}
\end{bmatrix},\]
 and 
 \[\ket{\lambda_{(j+30)}} = \frac{1}{\sqrt{2}}\begin{bmatrix}
    \ket{v_j}\\
    \mathcal{J} \ket{v_j}
\end{bmatrix},\] 
 
 for $1 \leq j \leq 30$.

Let us say $\ket{u_1} = [u_{(1,1)}, u_{(1,2)}, u_{(1,3)}, \dots, u_{(1,30)}]^T$, then 
\[ J \ket{u_1} = \begin{bmatrix}
    u_{(1,30)}\\
    u_{(1,29)} \\
    u_{(1,28)} \\
    \dots  \\
    u_{(1,1)}
\end{bmatrix}.\]

The same follows for the $\ket{v_j}$ eigenvector. This means that the $x$-th entry of $\ket{\lambda_j}$ for some $j$ is the same as the $(61-x)$-th entry of $\ket{\lambda_j}$ up to a sign, for $1 \leq x \leq 60$, i.e.

\[\langle x | \lambda_j\rangle = \pm \langle (61-x) | \lambda_j\rangle.\]

\section{Limiting distribution of continuous time quantum walk of fullerene graph: code}
In this code, the input is the number of vertices in the graph and the algorithm outputs the quantum walk limiting distribution on the graph.
\begin{lstlisting}[language=Python, caption=Limiting distribution code for fullerenes of order ten]
%Vertices of the fullerene graph as input:
N = ;

M = zeros(N,N);

%Generates Adjacency matrix of fullerene of order 10.

for j = 7 :2: N-16
    
    M(j,j+11)= 1;
    M(j+11,j)= 1;
end
    
for j = 1: 5
    
    M(j,4+2*j) = 1;
    M(4+2*j,j) = 1;
end

k = N-4;

for j = N-13 : 2 :N - 6
    
    k = k+1;
    
    M(j,k) = 1;
    M(k,j) = 1;
    
end

for j = 15: 10 : N
  
M(j, j+1) = 1;
M(j+1, j) = 1;

end
  
for m = 0 : (N/10)-1
    
    %{
    if m == 0 
        j =1;
    else
        j = (2*m-1)*5+1;
        
    end
    %}
 
for  j= m*5+1 : 10*m+5 

    if j < 10*m+5
       
    M(j,j+1) = 1;
    M(j+1,j) = 1;
    
    elseif j == 5
        M(j,5*m+1)= 1;
        M(5*m+1,j)= 1;
    else
        M(j,j-9)= 1;
        M(j-9,j)= 1;
        
    end
    
end

end

for j = N-4 : N
    
    if j < N
        M(j,j+1) = 1;
        M(j+1,j) = 1;
    else
        M(j,N-4) = 1;
        M(N-4,j) = 1;
    end
end

for b = 15: 10 : (N-25)
    
    M(b,b+11) = 0;
    M(b+11,b) = 0;
    
end

A = M;

%gives the eigenvalues and eigenvectors
[V,D]= eig(A);

%temporary storage for the limiting probability distribution
Am = zeros(1, N);
%each raw stores the limiting distribution starting at the row-indexed initial state
K = zeros(N, N);

%Calculating the distribution
for k = 1 : N 

for j = 1 : N
    
Pa = 0;
Amp = 0;

for i = 1 : N-1
    
    if abs(D(i,i)-D(i+1, i+1)) <= 10^(-6) 
        
        Amp= Amp + V(k,i)*V(j,i); 
    else
        Amp=Amp +V(k,i)*V(j,i);
        Pa = Pa + abs(Amp)^2;
        Amp = 0;
    end     
end

Pa = Pa+abs(V(k,N)*V(j,N))^2;

Am(1, j)= Pa;
end
K(k, :)= Am;
end

%3d plot of limiting distribution
surf(K)

title('fullerene-N Quantum walk limiting distribution')
xlabel('vertices')
ylabel('starting vertex')
zlabel('Averaged probability starting at different vertices')
\end{lstlisting}

\end{document}